\documentclass[preprint,12pt]{article} 

\usepackage{amssymb}
\usepackage{amsthm}
\usepackage{amsmath}
\usepackage{ytableau}
\usepackage{dsfont}
\usepackage{tikz}
\usepackage{multirow}
\usepackage{caption,subcaption}
\usepackage{hyperref}
\usepackage{colortbl}
\usepackage[symbol]{footmisc}
\usepackage{comment}

\theoremstyle{plain}
   
   \newtheorem{theorem}{Theorem}[section]

   \newtheorem{conjecture}[theorem]{Conjecture}
   \newtheorem{problem}[theorem]{Problem}
   
   \theoremstyle{definition}

   \newtheorem{example}[theorem]{Example}

   \newtheorem{remark}[theorem]{Remark}

\newcommand{\CC}{{\mathbb {C}}}

\newcommand{\ZZ}{{\mathbb {Z}}}

\newcommand{\SSYT}{{\rm SSYT}}

\newcommand{\ch}{{\operatorname{ch}}}

\newcommand{\act}{\mathop{\mathrm{act}}}

\newcommand\scalemath[2]{\scalebox{#1}{\mbox{\ensuremath{\displaystyle #2}}}}

\DeclareMathOperator{\Gr}{Gr}

\title{Clustering Cluster Algebras with Clusters}
\author{Man-Wai Cheung\textsuperscript{a}\footnote{manwai.cheung@ipmu.jp}, Pierre-Philippe Dechant\textsuperscript{b,c,d}\footnote{p.p.dechant@leeds.ac.uk}, Yang-Hui He\textsuperscript{e,f,g,h}\footnote{hey@maths.ox.ac.uk},\\ Elli Heyes\textsuperscript{f,e}\footnote{elli.heyes@city.ac.uk}, Edward Hirst\textsuperscript{f,e}\footnote{edward.hirst@city.ac.uk}, Jian-Rong Li\textsuperscript{i}\footnote{lijr07@gmail.com (corresponding)}}
\date{\today\\ \mbox{}\\ 
\textsuperscript{a}\textit{\small School of Mathematics, Kavli IPMU (WPI), UTIAS, \\ The University of Tokyo, Kashiwa, Japan, 277-8583}\\
\textsuperscript{b}\textit{\small School of Mathematics, University of Leeds, Leeds, LS2 9JT}\\
\textsuperscript{c}\textit{\small Department of Mathematics, University of York, York, YO10 5DD}\\
\textsuperscript{d}\textit{\small York Cross-disciplinary Centre for Systems Analysis, \\ University of York, York, YO10 5DD}\\
\textsuperscript{e}\textit{\small London Institute for Mathematical Sciences, \\ Royal Institution, London W1S 4BS, UK}\\
\textsuperscript{f}\textit{\small Department of Mathematics, City, University of London, EC1V 0HB, UK}\\
\textsuperscript{g}\textit{\small Merton College, University of Oxford, OX1 4JD, UK}\\
\textsuperscript{h}\textit{\small School of Physics, NanKai University, Tianjin, 300071, P.R. China}\\
\textsuperscript{i}\textit{\small Faculty of Mathematics, University of Vienna, \\ Oskar-Morgenstern-Platz 1, Vienna, 1090}
}

\begin{document}

\maketitle 
\newpage
\begin{abstract}
Classification of cluster variables in cluster algebras (in particular, Grassmannian cluster algebras) is an important problem, which has direct applications to computations of scattering amplitudes in physics. In this paper, we apply the tableaux method to classify cluster variables in Grassmannian cluster algebras $\CC[\Gr(k,n)]$ up to $(k,n)=(3,12), (4,10)$, or $(4,12)$ up to a certain number of columns of tableaux, using HPC clusters. These datasets are made available on GitHub. Supervised and unsupervised machine learning methods are used to analyse this data and identify structures associated to tableaux corresponding to cluster variables.    
Conjectures are raised associated to the enumeration of tableaux at each rank and the tableaux structure which creates a cluster variable, with the aid of machine learning.
\end{abstract}



\noindent \textit{Report Number:} LIMS-2022-025


\section{Introduction}
\label{sec:intro}

Cluster algebras were first introduced by Fomin and Zelevinsky in 2000 \cite{FZ02} in the context of Lie theory but have since been applied to many other areas of mathematics and physics, such as in integrable systems, tropical geometry, and scattering amplitudes. Classification of cluster variables (in particular cluster variables in Grassmannian cluster algebras) is important in mathematics \cite{HL10, JKS} and scattering amplitudes in physics \cite{ABCGPT, ALS, DDHMPS, DFGK, GGSVV, HP, Franco:2014nca, Franco:2017lpa, Franco:2003ja}. 

For example, in mathematics, cluster variables in Grassmannian cluster algebras $\CC[\Gr(k,n)]$ correspond to real prime modules of the quantum affine algebra $U_q(\widehat{\mathfrak{sl}_k})$, \cite{HL10, CDFL}. They also correspond to rigid indecomposable modules in Grassmannian cluster categories \cite{BBGL}. In physics, some particular scattering amplitudes in $N = 4$ super Yang-Mills theory can be written as sums of polylogarithms in variables with a cluster algebra structure, \cite{GGSVV}. Remainder functions of MHV scattering amplitudes in the planar limit of $N = 4$ super Yang-Mills theory tend to be linear combinations of generalized polylogarithms whose symbols are composed of $X$-coordinates of the the cluster algebra $\CC[\Gr(4,n)]$, \cite{GGSVV}. Cluster $X$-coordinates can be obtained from cluster $A$-coordinates. In this paper, cluster $A$-coordinates are called cluster variables.

In short, a cluster algebra is a commutative ring generated inside an ambient field. It is defined iteratively from an initial seed, consisting of a set of variables, called a cluster, and a quiver, via a procedure called mutation. The mutation process produces further seeds, which consist of clusters and quivers. The cluster algebra is the algebra generated by all cluster variables (including frozen variables); see §\ref{subsec:cluster algebras} for more details. 

As a set, for integers $k \le n$, the Grassmannian variety $\Gr(k,n)$ is the set of all $k$-dimensional subspaces of the $n$ dimensional vector space $\CC^n$. Scott \cite{Sco} proved that there is a cluster algebra structure in the coordinate ring $\CC[\Gr(k,n)]$. The ring $\CC[\Gr(k,n)]$ is called a Grassmannian cluster algebra.


Hernandez and Leclerc \cite{HL10} showed that there is a cluster algebra structure on the Grothendieck ring $K_0(\mathcal{C}_{\ell})$ of a certain subcategory $\mathcal{C}_{\ell}$ of the category of finite-dimensional modules of a quantum affine algebra $U_q(\widehat{\mathfrak{g}})$. In the case when $\mathfrak{g} = \mathfrak{sl}_k$, the cluster algebra $K_0(\mathcal{C}_{\ell})$ is isomorphic to the cluster algebra $\CC[\Gr(k,n,\sim)]$, where $\CC[\Gr(k,n,\sim)]$ is the quotient of the Grassmannian cluster algebra $\CC[\Gr(k,n)]$ by identifying certain frozen variables with $1$, cf. \cite{HL10, Sco}. 

Simple modules of $U_q(\widehat{\mathfrak{g}})$ are parametrized by dominant monomials in formal variables $Y_{i,s}$, $i \in I$, $s \in \mathbb{Z}$, where $I$ is the set of vertices of the Dynkin diagram of $\mathfrak{g}$, cf. \cite{CP94}. It is shown in \cite{CDFL} that, in the case of $\mathfrak{g} = \mathfrak{sl}_k$, the monoid of dominant monomials is isomorphic to the monoid $\SSYT(k, [n], \sim)$, where $\SSYT(k, [n], \sim)$ is a quotient of the monoid $\SSYT(k, [n])$ of semistandard tableaux (SSYT) of rectangular shape with $k$ rows and with entries in $[n] = \{1, \ldots, n\}$, cf. §\ref{subsec:grkn and semistandard Young tableaux}. Therefore, every simple module of $U_q(\widehat{\mathfrak{sl}_k})$ corresponds to a semistandard Young tableau in $\SSYT(k, [n], \sim)$. It follows, that the dual canonical basis of $\CC[\Gr(k,n,\sim)]$ is in one-to-one correspondence with tableaux in $\SSYT(k, [n], \sim)$. In particular, cluster variables in $\CC[\Gr(k,n,\sim)]$ correspond to a tableau in $\SSYT(k, [n], \sim)$.

The set of cluster variables in $\CC[\Gr(k,n)]$ is the union of the set of cluster variables in $\CC[\Gr(k,n,\sim)]$ and the frozen variables in $\CC[\Gr(k,n)]$. Therefore, in order to classify cluster variables in $\CC[\Gr(k,n)]$, it suffices to classify cluster variables in $\CC[\Gr(k,n,\sim)]$. We say that a tableau in $\SSYT(k, [n], \sim)$ (resp. $\SSYT(k, [n])$) is a cluster variable if the dual canonical basis element corresponding to it is a cluster variable in $\CC[\Gr(k,n,\sim)]$ (resp. $\CC[\Gr(k,n)]$). Up to frozen variables, cluster variables in $\SSYT(k, [n], \sim)$ and $\SSYT(k, [n])$ are the same. 

A simple $U_q(\widehat{\mathfrak{g}})$-module $M$ is called real if $M \otimes M$ is simple, cf. \cite{Le03}. A simple $U_q(\widehat{\mathfrak{g}})$-module $M$ is called prime if $M$ is not isomorphic to $M_1 \otimes M_2$ for any non-trivial modules $M_1, M_2$, cf. \cite{CP97}. Hernandez and Leclerc \cite{HL10} conjectured that real prime modules of $U_q(\widehat{\mathfrak{g}})$ are in one-to-one correspondence to cluster variables in $K_0(\mathcal{C}_{\ell})$. Therefore it is important to classify cluster variables in $K_0(\mathcal{C}_{\ell})$, and equivalently in $\CC[\Gr(k,n)]$.

In the context of planar $N=4$ super Yang-Mills theory, the cluster variables in $\CC[\Gr(k,n)]$ appear as symbol letters of scattering amplitudes, \cite{ALS, DDHMPS, DFGK, GGSVV, HP, HP2021}. Therefore classification of cluster variables in $\CC[\Gr(k,n)]$ is also important in physics. 

For $T \in \SSYT(k,[n])$, we say that $T$ is of rank $d$ if $T$ has $d$ columns. We say that a cluster monomial (in particular, a cluster variable) is of rank $r$ if the corresponding tableau has rank $r$. For general $k \le n$, the number of cluster variables in $\CC[\Gr(k,n)]$ is infinite. On the other hand, if we count cluster variables in $\CC[\Gr(k,n)]$ with a given rank, then the number is finite. This is because the number of semistandard Young tableaux with a given rank is finite. 

In this paper, we apply high-performance computing (HPC) clusters to compute cluster variables in $\CC[\Gr(k,n)]$. Our method of computing cluster variables is to use mutation of tableaux introduced in \cite{CDFL}, cf. Formula (\ref{eq:howtomutate}). 
We compute the cluster variables in the Grassmannian cluster algebras $\CC[\Gr(3,12)]$ up to rank 6, $\CC[\Gr(4,12)]$ up to rank 4, and $\CC[\Gr(4,10)]$ up to rank 6, cf. Table \ref{table:number of cluster variables in Grkn}. The \href{https://github.com/edhirst/GrassmanniansML.git}{datasets} produced amount to $\sim 0.75$Gb of data and took $\sim 0.5$ million core hours to compute. 

From our results, we obtain conjectural formulas for numbers of cluster variables of certain ranks in $\CC[\Gr(3,n)]$ and $\CC[\Gr(4,n)]$, cf. Conjecture \ref{conj:formula of number of cluster variables}:  
\begin{align*}
& N_{3,n,3} = 24 {n \choose 8} + 9 {n \choose 9}, \\
& N_{3,n,4} = 288 {n \choose 9} + 400 {n \choose 10} + 264 {n \choose 11} + 52 {n \choose 12}, \\
& N_{4,n,3} = 174 {n \choose 8} + 855 {n \choose 9} + 1285 {n \choose 10} + 693 {n \choose 11} + 123 {n \choose 12},
\end{align*} 
where $N_{k,n,r}$ is the number of cluster variables of rank $r$ in $\CC[\Gr(k,n)]$.

We also conjecture that when one replaces a set of numbers $a_1< \ldots < a_m$ appearing in a cluster variable (tableau) with another set of numbers $a_1'< \ldots < a_m'$, one will obtain another cluster variable, cf. Conjecture \ref{conj:expansion of a cluster variable is a cluster variable}. 

Grassmannian cluster algebras have many cluster variables, forming large datasets with rich structure. Therefore, in addition to computing this data and making it \href{https://github.com/edhirst/GrassmanniansML.git}{readily available} for physical and mathematical application, we also turn to techniques from data science and machine learning (ML) to analyse these datasets of variables and extract some of this structure.

More specifically, we would like to study the following problems:

\begin{problem}\label{ml_prob1}
Can machine learning methods identify whether a given semistandard Young tableau corresponds to a cluster variable?
\end{problem} 

\begin{problem}\label{ml_prob2}
What structure of these tableaux can be extracted by machine learning techniques which identifies the tableau as corresponding to a cluster variable?
\end{problem} 

The machine learning methods we employ include both supervised and unsupervised methods. Support Vector Machines and Neural Networks both learn to distinguish -- with strong performance -- tableaux from different algebras, and also learn to distinguish those tableaux that are cluster variables from those that are not.
Principal Component Analysis and K-Means Clustering, also highlight to us the key features in the tableau data. 

This paper is structured as follows: In §\ref{background} we establish the relevant mathematical background surrounding Grassmannian cluster algebra cluster variables and their representation as semistandard Young tableaux. In §\ref{Gr_CAs} we provide information regarding the computation of cluster variables in Grassmannian cluster algebras. In §\ref{ml} we analyse the generated data using techniques from supervised and unsupervised machine learning. Conclusions are presented in §\ref{conc}.

Coding scripts, and data used in this work are available at the respective GitHub repository: \url{https://github.com/edhirst/GrassmanniansML.git}

\section{Grassmannian cluster algebras}\label{background}
In this section, we recall results in \cite{FZ02, CDFL, Sco} about cluster algebras and Grassmannian cluster algebras.  


\subsection{Cluster algebras} \label{subsec:cluster algebras}
We begin by recalling the definition of cluster algebras given by Fomin and Zelevinsky \cite{FZ02}. 

A quiver $Q=(Q_0, Q_1, s, t)$ is a directed graph without loops or $2$-cycles that can be described by a vertex set $Q_0$, an arrow set $Q_1$, and maps $s,t: Q_1 \to Q_0$ that take an arrow to its source and target, respectively. We identify $Q_0 = [m] = \{1,\dots,m\}$ and declare vertices $1,\dots,r$ as mutable vertices and vertices $r+1,\dots,m$ as frozen vertices. 

For $k \in [r]$, the mutated quiver $\mu_k(Q)$ is a quiver obtained from $Q$ by:
\begin{enumerate}
\item[(i)] for each sub-quiver $i \to k \to j$, add a new arrow $i \to j$,

\item[(ii)] reverse the orientation of every arrow with target or source equal to $k$,

\item[(iii)] remove the arrows in a maximal set of pairwise disjoint $2$-cycles. 
\end{enumerate}

Let $\mathcal{F}$ be an ambient field abstractly isomorphic to a field of rational functions in $m$ independent variables. A seed in $\mathcal{F}$ is a pair $({\bf x}, Q)$, where ${\bf x} = (x_1, \ldots, x_m)$ is a free generating set of $\mathcal{F}$, called a cluster, and $Q$ is a quiver. The variables $x_1, \ldots, x_r$ are called cluster variables, and the variables $x_{r+1}, \ldots, x_m$ are called frozen variables. 

For a seed $({\bf x}, Q)$ and $k \in [r]$, the mutated seed $\mu_k({\bf x}, Q)$ is  $({\bf x}', \mu_k(Q))$, where ${\bf x}' = (x_1', \ldots, x_m')$ with $x_j'=x_j$ for $j\ne k$ and $x_k' \in \mathcal{F}$ determined by
\begin{align*}
x_k' x_k = \prod_{\alpha \in Q_1, s(\alpha)=k} x_{t(\alpha)} + \prod_{\alpha \in Q_1, t(\alpha)=k} x_{s(\alpha)}.
\end{align*} 

After making a choice of an initial labeled seed, we say that a seed is reachable if it can be obtained from the initial seed by a finite sequence of mutations. One defines the clusters (resp. cluster variables) to be the clusters (resp. cluster variables) appearing in all reachable seeds. Two cluster variables are called compatible if they appear together in a cluster. A cluster monomial is a product of compatible cluster variables. The cluster algebra is the $\mathbb{C}$-algebra generated by all cluster variables and frozen variables. 

\subsection{Grassmannian cluster algebras and semistandard Young tableaux} \label{subsec:grkn and semistandard Young tableaux}

We denote by $\Gr(k,n)$ the Grassmannian of $k$-planes in $\mathbb{C}^n$ and $\CC[\Gr(k,n)]$ its homogeneous coordinate ring. It was shown by Scott \cite{Sco} that the ring $\CC[\Gr(k,n)]$ has a cluster algebra structure. 
Furthermore, it was shown in \cite{CDFL} that every cluster monomial (in particular, every cluster variable) in $\CC[\Gr(k,n)]$ corresponds to a semistandard Young tableau.  
This was achieved by using the isomorphism between two cluster algebras: one cluster algebra is the Grothendieck ring of a certain subcategory of the category of finite-dimensional modules of the quantum affine algebra $U_q(\widehat{\mathfrak{sl}_k})$; the other cluster algebra is $\CC[\Gr(k,n,\sim)]$, where $\CC[\Gr(k,n,\sim)]$ is the quotient of $\CC[\Gr(k,n)]$ by the ideal $\langle P_{i,i+1, \ldots, i+k-1}-1, \quad i \in [n-k+1] \rangle$. 


For $k \le n$, we denote by $\SSYT(k,[n])$ the set of all semistandard Young tableaux of rectangular shape with $k$ rows and with entries in $[n] = \{1, \ldots, n\}$. 

For $S,T \in {\rm SSYT}(k, [n])$, we denote by $S \cup T$ the row-increasing tableau whose $i$th row is the union of the $i$th rows of $S$ and $T$ (as multisets). It was shown in \cite{CDFL} that for any $S,T \in {\rm SSYT}(k, [n])$, $S \cup T$ is in ${\rm SSYT}(k, [n])$. 

We call $S$ a factor of $T$, and write $S \subset T$, if the $i$th row of $S$ is contained in that of $T$ (as multisets), for $i \in [k]$. In this case, we define $\frac{T}{S}=S^{-1}T=TS^{-1}$ to be the row-increasing tableau whose $i$th row is obtained by removing that of $S$ from that of $T$ (as multisets), for $i \in [k]$. 

A tableau $T \in {\rm SSYT}(k, [n])$ is trivial if each entry of $T$ is one less than the entry below it. 

For any $T \in {\rm SSYT}(k, [n])$, we  denote by $T_{\text{red}} \subset T$ the semistandard tableau obtained by removing a maximal trivial factor from $T$. That is, $T_{\text{red}}$ is the tableau with the minimal number of columns such that $T = T_{\text{red}} \cup S$ for a trivial tableau $S$. For trivial $T$, one has $T_{\text{red}} = \mathds{1}$ (the empty tableau). For $S, T \in {\rm SSYT}(k, [n])$, we define $S \sim T$ if $S_{\text{red}} = T_{\text{red}}$. The reduction 
relation ``$\sim$'' is an equivalence relation. We denote by ${\rm SSYT}(k, [n],\sim)$ the set of $\sim$-equivalence classes.

We use the same notation for a tableau $T$ and its equivalence class, writing either $T \in {\rm SSYT}(k, [n])$ or $T \in {\rm SSYT}(k, [n],\sim)$ when it is important to distinguish these. 

\begin{example}
We illustrate the operations $\cup$ and $\sim$: 
\begin{align*}
\scalemath{0.7}{
\begin{ytableau}
1 & 3    \\
2 & 7 \\
6 & 11
\end{ytableau}
\cup
\begin{ytableau}
1 & 7   \\
2 & 9 \\
8 & 10
\end{ytableau} = 
\begin{ytableau}
1 & 1 & 3 & 7   \\
2 & 2 & 7 & 9 \\
6 & 8 & 10 & 11
\end{ytableau} }
\text{ and }
\scalemath{0.7}{
\begin{ytableau}
1 \\
3 \\
6 
\end{ytableau} \sim 
\begin{ytableau}
1 & 2 & 3  \\
3 & 3 & 4 \\
4 & 5 & 6
\end{ytableau}. }
\end{align*}
\end{example}

A one-column tableau is called a fundamental tableau if its content is $[i,i+k] \setminus \{r\}$ for $r \in \{i+1, \ldots, i+k-1\}$. A tableau $T$ is said to have small gaps if each of its columns is a fundamental tableau. Any tableau in $\SSYT(k,[n])$ is $\sim$-equivalent to a unique small gap tableau.

\subsection{Dominance order}

Let $\lambda = (\lambda_1,\dots,\lambda_\ell)$ with $\lambda_1 \geq \lambda_2 \ge \cdots \geq \lambda_\ell \geq 0$ and $\mu = (\mu_1,\dots,\mu_\ell)$ with $\mu_1 \ge \mu_2 \ge \cdots \ge \mu_{\ell} \ge 0$ be partitions. Then $\lambda \geq \mu$ in dominance order if $\sum_{j \leq i}\lambda_j \geq \sum_{j \leq i}\mu_j$ for $i=1,\dots,\ell$. For a tableau $T$ (, not necessarily rectangular shape), let ${\rm sh}(T)$ denote the shape of $T$. That is, ${\rm sh}(T) = (\lambda_1, \ldots, \lambda_r)$, where $\lambda_i$ is the number of boxes of $T$ in the $i$th row.  
For $i \in [n]$, let $T[i]$ denote the restriction of $T \in {\rm SSYT}(k,[n])$ to the entries in $[i]$. That is, $T[i]$ is the tableau obtained from $T$ by removing all boxes which have numbers greater than $i$. 

For a tableau $T$, we call the multi-set of numbers appearing (count multiplicities) in $T$ the content of $T$. 
For $T,T' \in {\rm SSYT}(k, [n])$ with the same content,
we say that $T \geq T'$ if ${\rm sh}(T[i]) \geq {\rm sh}(T'[i])$ in the dominance order on partitions, for $i=1,\dots,n$.  

\subsection{Cluster variables in $\CC[\Gr(k,n)]$}

Recall that $\CC[\Gr(k,n,\sim)]$ is the quotient of $\CC[\Gr(k,n)]$ by the ideal 
\begin{align*}
\langle P_{i,i+1, \ldots, i+k-1}-1, \quad i \in [n-k+1] \rangle.
\end{align*} 

Theorem 3.25 in \cite{CDFL} states that every cluster variable in the cluster algebra $\CC[\Gr(k,n,\sim)]$ is of the form $\ch(T)$ (see Equation (\ref{eq:formula of ch(T)}); the notation $\ch(T)$ is used because it corresponds to the $q$-character of a module of the quantum affine algebra $U_q(\widehat{\mathfrak{sl}_k})$) 
for some real prime $T \in {\rm SSYT}(k,[n])$ (a tableau is called real (resp. prime) if the corresponding quantum affine algebra module is real (resp. prime), \cite{CDFL}). An explicit formula of $\ch(T)$ is given in Theorem 5.8 of \cite{CDFL}:
\begin{align}\label{eq:formula of ch(T)}
\ch(T) = \sum_{u \in S_k} (-1)^{\ell(uw_T)} p_{uw_0, w_Tw_0}(1) P_{u; T'} \in \CC[\Gr(k,n,\sim)]
\end{align}
where $T'$ is the small gap tableau such that $T \sim T'$, $P_{u ; T'}$ is some monomial of Pl\"{u}cker coordinates, $w_T$ is some permutation in $S_k$, and $p_{u,v}(q)$ is a Kazhdan-Lusztig polynomial.  
When $\ch(T)$ is a cluster variable, we also call $T$ itself a cluster variable.  

\begin{remark}
The set of cluster variables in $\CC[\Gr(k,n)]$ is the union of the set of cluster variables in $\CC[\Gr(k,n,\sim)]$ and frozen variables in $\CC[\Gr(k,n)]$. The frozen variables (up to sign) in $\CC[\Gr(k,n)]$ are $P_{i,i+1, \ldots, i+k-1}$, $i \in [n]$, where $i+n$ are identified with $i$. The frozen variables correspond to one-column tableaux with consecutive entries or with entries $\{1,2,\ldots,r, n-k+r+1, \ldots, n-1, n\}$, $r \in [k-1]$. 
\end{remark}

We compute cluster variables in $\CC[\Gr(k,n)]$ in the following way \cite[Section 4]{CDFL}: Starting from the initial seed of $\CC[\Gr(k,n)]$, each time we perform a mutation at the cluster variable $\ch(T_r)$, we obtain a cluster variable $\ch(T'_r)$ defined recursively by 
\begin{align*}
\ch(T'_r)\ch(T_r) = \prod_{i \to r} \ch(T_i) + \prod_{r \to i} \ch(T_i),
\end{align*}
with $\ch(T_i)$ the cluster variable at the vertex $i$. Denote by $\max\{\cup_{i \to r} T_i, \cup_{r \to i} T_i \}$ the tableau which is larger in the dominance order. The tableau $T'$ corresponding to the new cluster variable $\ch(T_r')$ can be computed by the following formula:
\begin{align}
T'_r = T^{-1}_r \max\{\cup_{i \to r} T_i, \cup_{r \to i} T_i \}.\label{eq:howtomutate}
\end{align}

The following are some examples of mutations in $\CC[\Gr(3,8)]$:
\begin{align*}
& \scalemath{0.7}{
\ch(\begin{ytableau}
1  \\
3 \\
4 
\end{ytableau}) \ch(\begin{ytableau}
2  \\
3 \\
5 
\end{ytableau}) = \ch(\begin{ytableau}
1  \\
3 \\
5 
\end{ytableau}) \ch(\begin{ytableau}
2  \\
3 \\
4 
\end{ytableau}) + \ch(\begin{ytableau}
1  \\
2 \\
3 
\end{ytableau}) \ch(\begin{ytableau}
3  \\
4 \\
5 
\end{ytableau}), } \\
& \scalemath{0.7}{ \ch(\begin{ytableau}
2  \\
3 \\
8 
\end{ytableau}) \ch(\begin{ytableau}
1 & 3 & 4  \\
2 & 5 & 6 \\
4 & 7 & 8 
\end{ytableau}) = \ch(\begin{ytableau}
1  \\
2 \\
8 
\end{ytableau}) \ch(\begin{ytableau}
3 & 4  \\
5 & 6 \\
7 & 8
\end{ytableau}) \ch(\begin{ytableau}
2  \\
3 \\
4 
\end{ytableau}) + \ch(\begin{ytableau}
3  \\
4 \\
8 
\end{ytableau}) \ch(\begin{ytableau}
2 & 4  \\
5 & 6 \\
7 & 8 
\end{ytableau}) \ch(\begin{ytableau}
1  \\
2 \\
3 
\end{ytableau}). }
\end{align*} 

\section{Cluster variables in Grassmannian cluster algebras}\label{Gr_CAs}

In this section, we describe the result of our computations of cluster variables in $\CC[\Gr(k,n)]$, giving rise to our \href{https://github.com/edhirst/GrassmanniansML.git}{dataset}. 

We generate cluster variables by performing random mutations and following formula (\ref{eq:howtomutate}). When we compute cluster variables with ranks less than or equal to a given number $r$, if we see a cluster variable (tableau) with rank greater than $r$, we mutate at that vertex again so that the cluster variable with rank greater than $r$ does not appear. In this way, the cluster variables we generate are always with ranks less or equal to $r$. In practice, we perform sufficiently many mutations and when we see that no new cluster variables appear after $\sim 10000$ core hours on the HPC cluster (approximately 10\% of the total core time per run), we conjecture that we have obtained all cluster variables with ranks less than or equal to $r$.

\subsection{Some finite-type cluster algebras} 

The cluster algebra $\CC[\Gr(3,3)]$ has only one frozen variable 
\begin{align*}
\scalemath{0.7}{
\begin{ytableau}
1 \\ 2 \\ 3
\end{ytableau}},
\end{align*}
and no mutable cluster variables. 

\

There are $4$ frozen variables in $\CC[\Gr(3,4)]$:
\begin{align*}
\scalemath{0.7}{
\begin{ytableau}
1 \\ 2 \\ 3
\end{ytableau}, \
\begin{ytableau}
1 \\ 2 \\ 4
\end{ytableau}, \
\begin{ytableau}
1 \\3 \\ 4
\end{ytableau}, \
\begin{ytableau}
2 \\ 3 \\ 4
\end{ytableau}}
\end{align*}
and no mutable cluster variables. 

\

In $\CC[\Gr(3,5)]$, there are $5$ cluster variables:
\begin{align*}
\scalemath{0.7}{
\begin{ytableau}
1 \\ 3 \\ 5
\end{ytableau}, \
\begin{ytableau}
2 \\ 3 \\ 5
\end{ytableau}, \
\begin{ytableau}
2 \\ 4 \\ 5
\end{ytableau}, \
\begin{ytableau}
1 \\ 2 \\ 4
\end{ytableau}, \
\begin{ytableau}
1 \\ 3 \\ 4
\end{ytableau}},
\end{align*}
and $5$ frozen variables:
\begin{align*}
\scalemath{0.7}{
\begin{ytableau}
1 \\ 2 \\ 3
\end{ytableau}, \
\begin{ytableau}
2 \\ 3 \\ 4
\end{ytableau}, \
\begin{ytableau}
3 \\ 4 \\ 5
\end{ytableau}, \
\begin{ytableau}
1 \\ 2 \\ 5
\end{ytableau}, \
\begin{ytableau}
1 \\ 4 \\ 5
\end{ytableau}}.
\end{align*}

\

In $\CC[\Gr(3,6)]$, there are $16$ cluster variables:
\begin{align*}
\scalemath{0.7}{
\begin{ytableau}
1 \\ 4 \\ 6
\end{ytableau}, \
\begin{ytableau}
3 \\ 4 \\ 6
\end{ytableau}, \
\begin{ytableau}
2 \\ 4 \\ 6
\end{ytableau}, \
\begin{ytableau}
2 \\ 3 \\ 6
\end{ytableau}, \
\begin{ytableau}
2 \\ 5 \\ 6
\end{ytableau}, \
\begin{ytableau}
2 \\ 4 \\ 5
\end{ytableau}, \
\begin{ytableau}
2 \\ 3 \\ 5
\end{ytableau}, \
\begin{ytableau}
1 \\ 2 \\ 5
\end{ytableau}, \
\begin{ytableau}
1 \\ 4 \\ 5
\end{ytableau}, \
\begin{ytableau}
1 \\ 3 \\ 5
\end{ytableau}, \
\begin{ytableau}
1 \\ 3 \\ 4
\end{ytableau}, \
\begin{ytableau}
1 \\ 3 \\ 6
\end{ytableau}, \
\begin{ytableau}
3 \\ 5 \\ 6
\end{ytableau}, \
\begin{ytableau}
1 \\ 2 \\ 4
\end{ytableau}, \
\begin{ytableau}
1 & 3 \\ 2 & 5 \\ 4 & 6
\end{ytableau}, \
\begin{ytableau}
1 & 2 \\ 3 & 4 \\ 5 & 6
\end{ytableau}, }
\end{align*}
and $6$ frozen variables: 
\begin{align*}
\scalemath{0.7}{
\begin{ytableau}
1 \\ 2 \\ 3
\end{ytableau}, \
\begin{ytableau}
2 \\ 3 \\ 4
\end{ytableau}, \
\begin{ytableau}
3 \\ 4 \\ 5
\end{ytableau}, \
\begin{ytableau}
4 \\ 5 \\ 6
\end{ytableau}, \
\begin{ytableau}
1 \\ 5 \\ 6
\end{ytableau}, \
\begin{ytableau}
1 \\ 2 \\ 6
\end{ytableau}.
}
\end{align*}

\

In $\CC[\Gr(3,7)]$, there are $28$ one-column tableaux which are cluster variables and $7$ one-column tableaux which are frozen variables. There are $14$ rank $2$ cluster variables which are obtained by sending $i \mapsto a_i$ in 
\begin{align*}
\scalemath{0.7}{
\begin{ytableau}
1 & 3 \\ 2 & 5 \\ 4 & 6
\end{ytableau}, \ 
\begin{ytableau}
1 & 2 \\ 3 & 4 \\ 5 & 6
\end{ytableau}},
\end{align*} 
where $a_1 < \cdots < a_6 \in \{1, \ldots, 7\}$.

\

In $\CC[\Gr(3,8)]$, there are $48$ one-column tableaux which are cluster variables and $8$ one-column tableaux which are frozen variables. There are $56$ rank $2$ cluster variables which are obtained by sending $i \mapsto a_i$ in the tableaux
\begin{align*}
\scalemath{0.7}{
\begin{ytableau}
1 & 3 \\ 2 & 5 \\ 4 & 6
\end{ytableau}, \
\begin{ytableau}
1 & 2 \\ 3 & 4 \\ 5 & 6
\end{ytableau}},
\end{align*}
where $a_1 < \cdots < a_6 \in \{1, \ldots, 8\}$. There are $24$ rank $3$ cluster variables:
\begin{align*} 
& \scalemath{0.7}{ \begin{ytableau} 1 & 3 & 4\\ 2 & 5 & 6\\ 4 & 7 & 8 \end{ytableau}, \ \begin{ytableau} 1 & 2 & 4\\ 2 & 3 & 7\\ 5 & 6 & 8 \end{ytableau}, \ \begin{ytableau} 1 & 2 & 3\\ 4 & 5 & 6\\ 6 & 7 & 8 \end{ytableau}, \ \begin{ytableau} 1 & 1 & 3\\ 2 & 5 & 6\\ 4 & 7 & 8 \end{ytableau}, \ \begin{ytableau} 1 & 3 & 4\\ 2 & 6 & 7\\ 5 & 7 & 8 \end{ytableau}, \ \begin{ytableau} 1 & 2 & 3\\ 3 & 4 & 5\\ 6 & 7 & 8 \end{ytableau}, \ \begin{ytableau} 1 & 2 & 5\\ 3 & 4 & 7\\ 6 & 8 & 8 \end{ytableau}, \ \begin{ytableau} 1 & 2 & 5\\ 3 & 4 & 7\\ 5 & 6 & 8 \end{ytableau}, } \\
& \scalemath{0.7}{ \begin{ytableau} 1 & 2 & 3\\ 4 & 4 & 5\\ 6 & 7 & 8 \end{ytableau}, \ \begin{ytableau} 1 & 3 & 4\\ 2 & 6 & 7\\ 5 & 8 & 8 \end{ytableau}, \ \begin{ytableau} 1 & 3 & 4\\ 2 & 5 & 6\\ 5 & 7 & 8 \end{ytableau}, \ \begin{ytableau} 1 & 2 & 5\\ 3 & 4 & 7\\ 6 & 6 & 8 \end{ytableau}, \ \begin{ytableau} 1 & 2 & 3\\ 2 & 5 & 6\\ 4 & 7 & 8 \end{ytableau}, \ \begin{ytableau} 1 & 2 & 3\\ 4 & 5 & 6\\ 7 & 7 & 8 \end{ytableau}, \ \begin{ytableau} 1 & 1 & 2\\ 3 & 4 & 5\\ 6 & 7 & 8 \end{ytableau}, \ \begin{ytableau} 1 & 2 & 4\\ 3 & 3 & 7\\ 5 & 6 & 8 \end{ytableau}, } \\
& \scalemath{0.7}{ \begin{ytableau} 1 & 2 & 4\\ 3 & 4 & 7\\ 5 & 6 & 8 \end{ytableau}, \ \begin{ytableau} 1 & 2 & 5\\ 3 & 4 & 7\\ 6 & 7 & 8 \end{ytableau}, \ \begin{ytableau} 1 & 2 & 3\\ 4 & 5 & 5\\ 6 & 7 & 8 \end{ytableau}, \ \begin{ytableau} 1 & 2 & 2\\ 3 & 4 & 5\\ 6 & 7 & 8 \end{ytableau}, \ \begin{ytableau} 1 & 3 & 4\\ 2 & 6 & 6\\ 5 & 7 & 8 \end{ytableau}, \ \begin{ytableau} 1 & 2 & 3\\ 4 & 5 & 6\\ 7 & 8 & 8 \end{ytableau}, \ \begin{ytableau} 1 & 1 & 4\\ 2 & 3 & 7\\ 5 & 6 & 8 \end{ytableau}, \ \begin{ytableau} 1 & 3 & 3\\ 2 & 5 & 6\\ 4 & 7 & 8 \end{ytableau}. }
\end{align*}  

\

For general $k \le n$, there are infinitely many cluster variables in $\CC[\Gr(k,n)]$. On the other hand, there are finitely many cluster variables in $\CC[\Gr(k,n)]$ with a fixed rank. For example, there are $168$ rank $2$ cluster variables and $225$ rank $3$ cluster variables in $\CC[\Gr(3,9)]$, \cite{BBGL}. 

As a core part of the work in this project, we have computed databases which contain all cluster variables with certain ranks in $\CC[\Gr(k,n)]$ for selected $k$, $n$ (noting these naturally contain all lower $n$s). Specifically, we compute the cluster variables as semistandard Young tableaux in the Grassmannian cluster algebras $\CC[\Gr(3,12)]$ up to rank 6, $\CC[\Gr(4,12)]$ up to rank 4, and $\CC[\Gr(4,10)]$ up to rank 6, totaling 2656212, 3089105, and 6346878 tableaux respectively. All datasets are available on \href{https://github.com/edhirst/GrassmanniansML.git}{GitHub}.

\subsection{Numbers of cluster variables with given ranks}

We denote by $N_{k,n,r}$, the number of cluster variables (including frozen variables) of rank $r$ in $\CC[\Gr(k,n)]$. Using high-performance computing, we stochastically compute all cluster variables with certain rank in $\CC[\Gr(k,n)]$. Since the process is stochastic we cannot explicitly verify this is an exhaustive list, however we note that in each case the last $\sim$10\% of runs did not generate any new variables. Therefore, the numbers $N_{k,n,r}$ in Table \ref{table:number of cluster variables in Grkn} provide at the very least lower bounds on the true number of cluster variables, and likely, equality. 

\begin{table}[!t]
\centering
\addtolength{\leftskip} {-2cm}
\addtolength{\rightskip}{-2cm} 
\scalebox{0.8}{
\begin{tabular}{|c|c|c|c|c|c|c|c|c|c|c|}
\hline
$r$ &  1 & 2& 3& 4& 5& 6& 7& 8 & 9 & 10\\
\hline
$N_{3,3,r}$ & 1 & 0 & 0 & 0&0 &0 &0 &0 &0 & 0 \\
\hline
$N_{3,4,r}$ & 4 & 0 & 0 & 0&0 &0 &0 &0 &0 & 0 \\
\hline
$N_{3,5,r}$ & 10 & 0 & 0 & 0&0 &0 &0 &0 &0 & 0 \\
\hline
$N_{3,6,r}$ & 20 & 2 & 0 & 0&0 &0 &0 &0 &0 & 0 \\
\hline
$N_{3,7,r}$ & 35 & 14 & 0 & 0&0 &0 &0 &0 &0 & 0 \\
\hline
$N_{3,8,r}$ & 56 & 56 & 24 & 0&0 &0 &0 &0 &0 & 0 \\
\hline
$N_{3,9,r}$ & 84 & 168 & 225 & 288 & 372 & 414 & 522 & 594 & 612 & 744  \\
\hline
$N_{3,10,r}$ & 120 & 420 & 1170 & 3280 & 8200 & 19140 & & & &  \\
\hline
$N_{3,11,r}$ & 165  & 924  & 4455  &  20504  & 77957 & 256553 & & & &  \\
\hline
$N_{3,12,r}$  &220   & 1848   & 13860  & 92980 & 486172 & 2061132 & & & &  \\
\hline 
\hline
$N_{4,4,r}$ &  1 & 0  & 0 & 0&0 &0 &0 &0 &0 & 0 \\
\hline
$N_{4,5,r}$ &  5 & 0  & 0 & 0&0 &0 &0 &0 &0 & 0 \\
\hline
$N_{4,6,r}$ &  15 & 0  & 0 & 0&0 &0 &0 &0 &0 & 0 \\
\hline
$N_{4,7,r}$ & 35  &  14 & 0 & 0&0 &0 &0 &0 &0 & 0 \\
\hline
$N_{4,8,r}$ & 70 & 120 & 174 & 208 & 296 & 304 & 420 & 416 & 536& 480  \\
\hline
$N_{4,9,r}$ & 126 & 576 & 2421 & 8622  & 27054  &  69390 &  &  & &   \\
\hline
$N_{4,10,r}$ & 210 & 2040 & 17665  & 117930 & 597500 & 2353760 & & & &  \\  
\hline 
$N_{4,11,r}$ & 330 & 5940  & 90563 & 980100 &  &  &  &  & &   \\
\hline 
$N_{4,12,r}$ & 495 & 15048  & 367479 & 5963856 &  &  &  &  & &   \\
\hline
\end{tabular} }
\caption{Number of cluster variables in $\CC[\Gr(k,n)]$ of rank $r$. Note each $N_{n,k,r}$ contains all those in $N_{n,k-1,r}$ by definition, so there are $(N_{n,k,r}-N_{n,k-1,r})$ new SSYT cluster variables for each box. Empty box entries denote variables to be computed in future work, and were beyond reasonable means for current computation.}
\label{table:number of cluster variables in Grkn}
\end{table}

It was proved in \cite{BBGL}, that the number of rank $2$ cluster variables in $\CC[\Gr(k,n)]$ ($k \le n/2$) is at least
\begin{align} \label{eq:number of rank 2 cluster variables}
N_{k,n,2} = \sum_{r=3}^{k} \left( \frac{2r}{3} \cdot p_1(r) + 2r \cdot p_2(r) + 4r \cdot p_3(r) \right) \cdot {n \choose 2r} {n-2r \choose k-r},
\end{align}
where $p_i(r)$ is the number of partitions $r=r_1+r_2+r_3$ such that $r_1,r_2,r_3 \in \ZZ_{\ge 1}$ and
$|\{r_1,r_2,r_3\}|=i$. According to our computation results, we expect that the number of rank $2$ cluster variables in $\CC[\Gr(k,n)]$ ($k \le n/2$) is exactly given by formula (\ref{eq:number of rank 2 cluster variables}). 

According to our computation results, we also have the following conjectures:
\begin{conjecture} \label{conj:formula of number of cluster variables}
The corresponding number of cluster variables is given by the following expressions
\begin{align*}
& N_{3,n,3} = 24 {n \choose 8} + 9 {n \choose 9}, \\
& N_{3,n,4} = 288 {n \choose 9} + 400 {n \choose 10} + 264 {n \choose 11} + 48 {n \choose 12}, \\
& N_{4,n,3} = 174 {n \choose 8} + 855 {n \choose 9} + 1285 {n \choose 10} + 693 {n \choose 11} + 123 {n \choose 12}. 
\end{align*}
\end{conjecture}

\begin{conjecture} \label{conj:expansion of a cluster variable is a cluster variable}
For any tableau $T \in {\rm SSYT}(k, [n])$ with entries $a_1 < \ldots < a_r$ and any function $f: \{a_1, \ldots, a_r\} \to [n']$, $n' \ge n$, such that $f(a_1) < \ldots < f(a_r)$, we have that $T$ is a cluster variable in $\CC[\Gr(k,n)]$ if and only if $f(T)$ is a cluster variable in $\CC[\Gr(k,n')]$. 
\end{conjecture}

\section{Machine Learning}\label{ml}

With the increase in computational power over the preceding decades, the ability to generate large amounts of mathematical data has become far more manageable. Where past mathematical work has focused on conjecture formulation from computation by hand on smaller samples of selected examples, now with larger datasets and algorithms to perform analysis computationally, data science is steadily becoming a key player in mathematical research.

Machine learning (ML), an umbrella field encompassing a large range of techniques from supervised, unsupervised, and reinforcement learning, has already seen a large amount of success in mathematics and related areas \cite{He:2017aed, Carifio:2017bov, Krefl:2017yox, Ruehle:2017mzq, Jejjala:2020wcc, Berglund:2021ztg, Cole:2021nnt, Arias-Tamargo:2022qgb, Berman:2021mcw, Bao:2021ofk, Bao:2021olg, Bao:2021auj, Hirst:2022qqr, He:2020eva, Manko:2022zfz, Chen:2022jwd}. Of particular relevance here, is past work on the use of ML to examine exchange graphs describing cluster seed interrelations \cite{Dechant:2022ccf,Bao:2020nbi}, built on \cite{musiker2011compendium}.

With the 3 large \href{https://github.com/edhirst/GrassmanniansML.git}{datasets} of cluster variables represented as SSYT, specifically $\{\CC[\Gr(3,12)]$ $r6$, $\CC[\Gr(4,10)]$ $r6$, $\CC[\Gr(4,12)]$ $r4\}$ where $r\#$ denotes the maximum rank (number of tableau columns) considered, we now apply a variety of techniques from ML to analyse them. 

\subsection{Data Formatting}
To ensure consistent formatting for data processing, all the SSYT were formatted as \texttt{numpy} arrays in \texttt{python}, padded with zeros up to the maximum size, such that they all had shape (4,6). 

\begin{figure}[htbp]
    \centering
    \begin{subfigure}{0.3\textwidth}
   	\centering
	\includegraphics[width=0.98\textwidth]{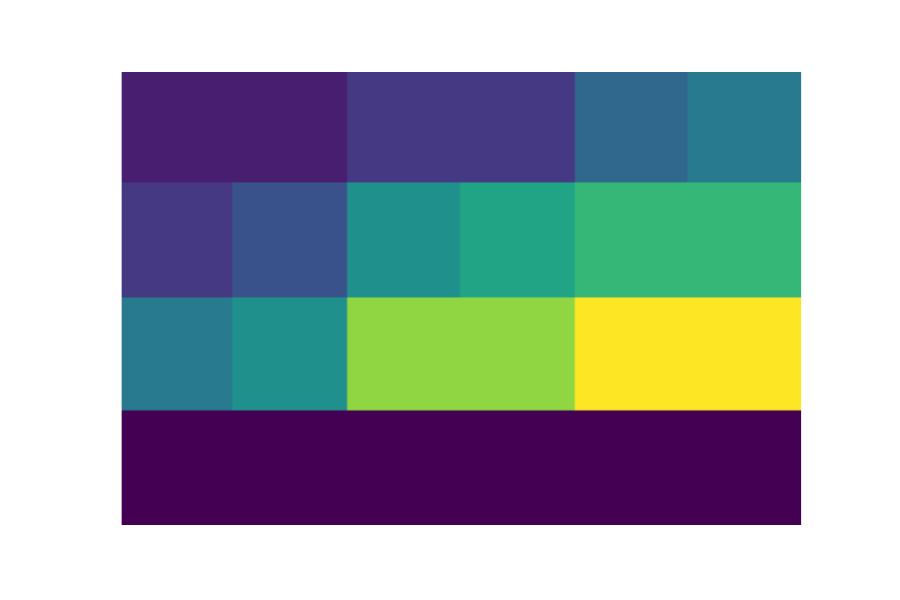}
        \caption{$\CC[\Gr(3,12)]$, $r \leq 6$}\label{312_eg}
    \end{subfigure}
    \begin{subfigure}{0.3\textwidth}
        \centering
        \includegraphics[width=0.98\textwidth]{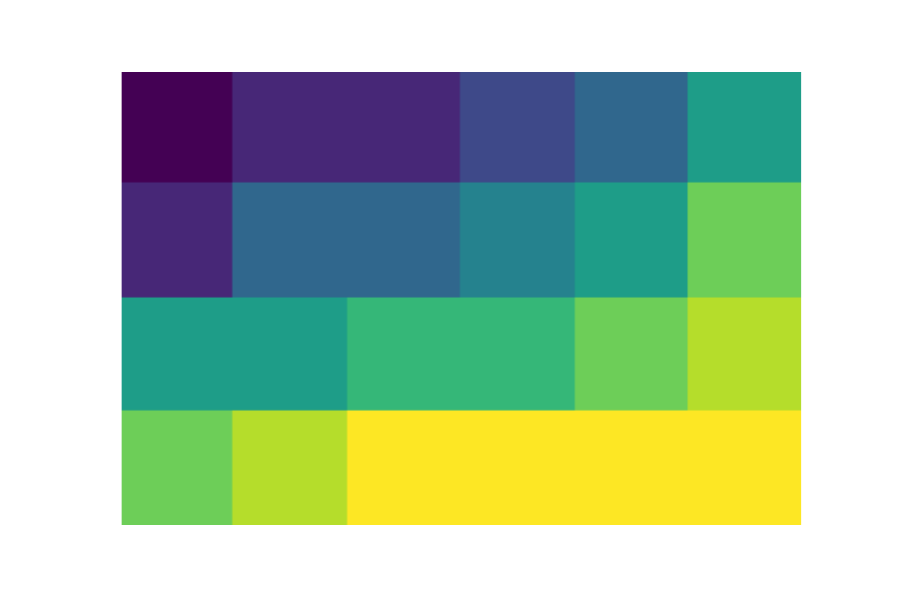}
        \caption{$\CC[\Gr(4,10)]$, $r \leq 6$}\label{410_eg}
    \end{subfigure} 
    \begin{subfigure}{0.3\textwidth}
        \centering
        \includegraphics[width=0.98\textwidth]{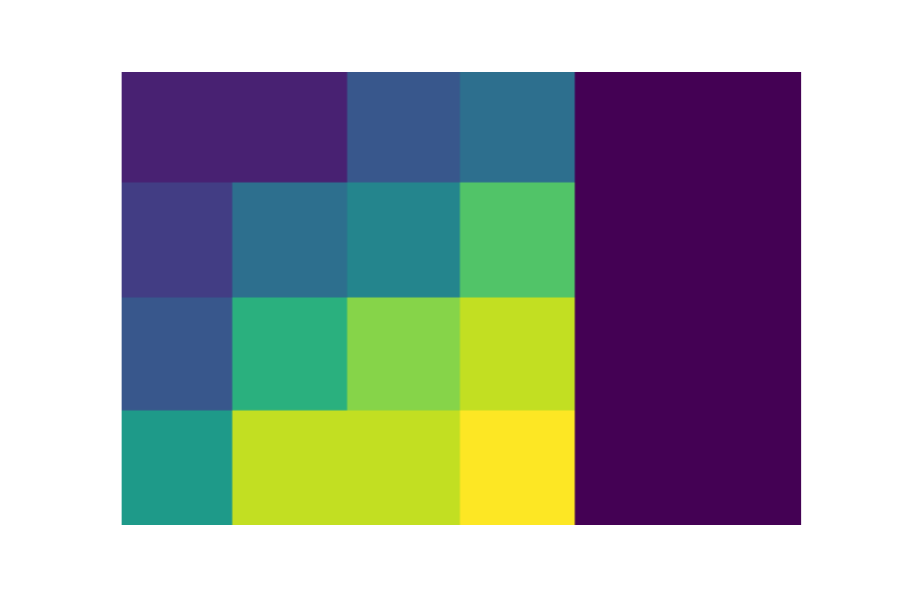}
        \caption{$\CC[\Gr(4,12)]$, $r \leq 4$}\label{412_eg}
    \end{subfigure}
\caption{Example images produced from the padded versions of the SSYT representing cluster variables in the respective Grassmannians. Note that for $\CC[\Gr(k,n)]$ $k$ represents the number of rows, $n$ the maximum entry, and $r$ the rank and hence the number of columns. These example images have the maximum rank in each case. }\label{eg_images}
\end{figure}

Examples of these arrays represented as images are shown in Figure \ref{eg_images}, where the clear padding of the $k=3$ cases (bottom row all zeros) and $r=4$ cases (right two columns all zeros) are shown for the $\CC[\Gr(3,12)]$ $r6$ and $\CC[\Gr(4,12)]$ $r4$ datasets respectively. The lighter box in the $\CC[\Gr(4,12)]$ $r4$ example indicates the higher maximum entry than $\CC[\Gr(4,10)]$ $r6$ as $n=12>10$.

These images just represent single examples from these Grassmannian cluster algebras. In each case lower rank SSYT are included (with more columns padded), as well as those with a smaller range of entries (where the colours are darker throughout the image).

\subsubsection{NCV Data Generation}
A point of key importance is that not all SSYT represent cluster variables in Grassmannian cluster algebras. 
Therefore one can create tableaux that increase along the $k$ rows, strictly increase down the $r$ columns, and with maximum entry $n$, which do not correspond to a cluster variable in the Grassmannian $\CC[\Gr(k,n)]$ with rank up to $r$.

Therefore, we refer back to Problem \ref{ml_prob1}: can ML techniques discover a relation that allows them to identify this? However, prior to applying ML methods, non-cluster-variable data, which we denote `NCV', corresponding to SSYT which are not cluster variables must be generated. Conversely, we denote the dataset of SSYT which are cluster variables as `CV'.

For each Grassmannian CV dataset an equivalent NCV dataset was generated, such that the number of rows, maximum entry, and maximum number of columns matched the Grassmannian's $k,n,r$ respectively.
Before generating each tableau, the number of columns was sampled from $[1,r]$. 
The NCV SSYTs were then initialised as random arrays of entries in the range $[1,n]$, which were sorted (enforcing an increase along the rows), then columns were checked and entries regenerated until the condition of strictly increasing down columns was met. 
Exhaustive checks were then applied to ensure each NCV SSYT was: 1) not in the respective Grassmannian CV dataset; 2) not already generated. These exhaustive checks ensure that the data is truly an NCV SSYT. We generate 10,000 tableaux for each NCV dataset, to be compared with a random sample of 10,000 tableaux from each respective Grassmannian.

\subsection{Supervised Classification}
In this subsection the content of Problem \ref{ml_prob1} is addressed, namely: how well can ML architectures learn to distinguish SSYT from different Grassmannian cluster algebras, or SSYT which are cluster variables from those which are not?

The ML architectures considered in this work are Support Vector Machines (SVM) and dense feed-forward Neural Networks (NN).

The goal of SVMs is to find $p-1$-dimensional hypersurfaces that best separate data points of different classes in $\mathbb{R}^p$, where in our case $p=24$ for the 24 entries of the SSYT. The shape of the hypersurface is dictated by the kernel style, and a regularisation parameter adds a cost to each parameter used to define the hypersurface -- discouraging them from overfitting and becoming too complicated. The hypersurfaces are fitted using training data and later  performance-tested with test data. The SVM we use here has regularisation parameter of 1.0 uses a Gaussian `rbf' kernel. We train until we reach a tolerance of 0.001 for fractional improvement in the proportion of correct classifications.

NNs are designed for complex non-linear function fitting. They are built out of perceptrons which take a vector as input, then output a number via linear action followed by non-linear activation: output $= \act \big( \sum_i (w_i \cdot \text{input}_i)+b\big)$ for weights $w_i$ and bias $b$. Layers of these perceptrons all connected to the subsequent layer make the NN dense, and feed-forward, as data flows through the network from initial input, through the layers, to final output.
The optimiser algorithm updates the weights and biases (by amounts proportional to the learning rate) during training to minimise the loss function, which is a measure of the difference between the NN predicted and the real output for each specific input, over batches of input data.
The NN architecture we use consists of three layers of size 16, 32 and 16, the perceptrons in each layer use ReLU activation, and the network is trained to minimise log-loss using the Adam optimisation algorithm, with batch size 200 and a learning rate of 0.001 until convergence below a tolerance of 0.0001. 

Both these architectures take the \texttt{sklearn} default hyperparameter values \cite{sklearn}. 

The learning performance is measured using accuracy and Matthew's correlation coefficient (MCC) on the test set predictions. These metrics may be described as functions on the confusion matrix (CM). Accuracy is the proportion of predictions that are correctly classified (i.e., the normalised sum of the diagonal of the confusion matrix). MCC is an analogue of this that accounts for off-diagonal terms, such that dataset bias is avoided in altering the validity of the measure.
The learning is carried out using 5-fold cross-validation, meaning that we train and test the network on 5 different partitions of the data, such that the union of the test sets equals the full dataset. This produces a set of 5 results for each learning measure, from which we compute an average and the standard error.

The first investigation uses NNs to perform multiclassification between the SSYT of the 3 Grassmannian CV datasets, whilst the second uses both SVMs and NNs to perform binary classification between the CV and NCV SSYT for each dataset. Due to the computational demands of training, random samples of 10,000 tableaux were taken from each Grassmannian CV dataset, to match the sizes of the generated NCV datasets.

\subsubsection{Grassmannian Multiclassification}\label{sec:multi_clf}
Before performing the NN supervised multiclassification between the 3 Grassmannian CV databases, it is noted that the $\CC[\Gr(4,10)]$ $r6$ and $\CC[\Gr(4,12)]$ $r4$ databases have natural overlap of tableaux in $\CC[\Gr(4,10)]$ $r4$. 
Using Table \ref{table:number of cluster variables in Grkn}, the two Grassmannians have 137845 variables in common; hence 137845/3089105 $\sim$ 0.04 of the $\CC[\Gr(4,10)]$ $r6$ data is in $\CC[\Gr(4,12)]$ $r4$ and 137845/6346878 $\sim$ 0.02 of the $\CC[\Gr(4,12)]$ $r4$ data is in $\CC[\Gr(4,10)]$ $r6$.
Therefore to avoid multi-labelling of tableaux for this classification problem the $\CC[\Gr(4,10)]$ $r4$ tableaux were removed from both relevant datasets. 

The learning measures, with standard error over the cross-validation, to 3 decimal places were:
\begin{align}
    Accuracy & = 1.000 \pm 0.000 \;,\\
    MCC & = 1.000 \pm 0.000 \;,\\
    CM & = \begin{pmatrix} 0.333 \pm 0.000 & 0.000 \pm 0.000 & 0.000 \pm 0.000 \\ 0.000 \pm 0.000 & 0.333 \pm 0.000 & 0.000 \pm 0.000 \\ 0.000 \pm 0.000 & 0.000 \pm 0.000 & 0.333 \pm 0.000 \end{pmatrix}\;,
\end{align}
where in the confusion matrix entries $CM_{ij}$ have $i$ index as the true class and $j$ index as the predicted class. The three Grassmannian classes are $(1,2,3) = (\CC[\Gr(3,12)] r6, \CC[\Gr(4,10)] r6, \CC[\Gr(4,12)] r4)$ respectively.

These results show perfect performance in identifying the Grassmannian a tableau belongs to. This is reassuring behaviour as already by eye one can distinguish the $\CC[\Gr(3,12)]$ tableaux by the number of rows, as well as the tableaux from the two $k=4$ databases due to the number of columns (i.e. rank) or maximum entry.

\subsubsection{Binary Classification of Cluster Variables from SSYT}
Learning measures for both SVM and NN architectures performing binary classification between the Grassmannian CV data and respective NCV data are shown in Table \ref{tab_rf}.

\begin{table}[htbp]
\begin{tabular}{|c|c|ccc|}
\hline
\multirow{2}{*}{Architecture} & \multirow{2}{*}{\begin{tabular}[c]{@{}c@{}}Learning\\ Measure\end{tabular}} & \multicolumn{3}{c|}{Grassmannian}  \\ \cline{3-5} & & \multicolumn{1}{c|}{$\mathbb{C}$[Gr(3,12)] $r6$} & \multicolumn{1}{c|}{$\mathbb{C}$[Gr(4,10)] $r6$} & $\mathbb{C}$[Gr(4,12)] $r4$ \\ \hline
\multirow{2}{*}{SVM} 
& Accuracy & \multicolumn{1}{c|}{\begin{tabular}[c]{@{}c@{}}0.913 \\ $\pm$ 0.002\end{tabular}} & \multicolumn{1}{c|}{\begin{tabular}[c]{@{}c@{}}0.928\\ $\pm$ 0.001\end{tabular}} & \begin{tabular}[c]{@{}c@{}}0.925\\ $\pm$ 0.001\end{tabular} \\ \cline{2-5} 
& MCC & \multicolumn{1}{c|}{\begin{tabular}[c]{@{}c@{}}0.830\\ $\pm$ 0.004\end{tabular}} & \multicolumn{1}{c|}{\begin{tabular}[c]{@{}c@{}}0.867\\ $\pm$ 0.004\end{tabular}} & \begin{tabular}[c]{@{}c@{}}0.852\\ $\pm$ 0.002\end{tabular} \\ \hline
\multirow{2}{*}{NN} 
& Accuracy & \multicolumn{1}{c|}{\begin{tabular}[c]{@{}c@{}}0.938\\ $\pm$ 0.002\end{tabular}} & \multicolumn{1}{c|}{\begin{tabular}[c]{@{}c@{}}0.946\\ $\pm$ 0.002\end{tabular}} & \begin{tabular}[c]{@{}c@{}}0.941\\ $\pm$ 0.002\end{tabular} \\ \cline{2-5} 
& MCC & \multicolumn{1}{c|}{\begin{tabular}[c]{@{}c@{}}0.878\\ $\pm$ 0.003\end{tabular}} & \multicolumn{1}{c|}{\begin{tabular}[c]{@{}c@{}}0.893\\ $\pm$ 0.005\end{tabular}} & \begin{tabular}[c]{@{}c@{}}0.885\\ $\pm$ 0.004\end{tabular} \\ \hline
\end{tabular}
\caption{Supervised binary classification between CV SSYT representing cluster variables in the respective Grassmannians, and NCV generated matrices designed to mimic them.}\label{tab_rf}
\end{table}

Both architecture styles are incredibly successful at determining the cluster variables from the full sets of SSYT. 
However, the still exceptional performance of the SVMs indicates that there is likely some unknown implicit structure in the SSYT entries that make them cluster variables.

In each case the NN architecture performs better, as may be expected since the architecture is more general.
The MCC scores correlate with accuracy, which is reassuring that the data is representative and unbiased, whilst the better performance for $\CC[\Gr(4,10)]$ $r6$ over $\CC[\Gr(4,12)]$ $r4$ implies that rank is a more important feature for determining the cluster variable property.
Explicit analysis of the misclassified SSYT in each case show no discernible pattern in the tableaux, confirming that the architectures are picking up on a more subtle structure for their learning.

These results confidently answer Problem \ref{ml_prob1} affirmatively: ML \textit{can} pick up on the underlying structure that makes a SSYT a cluster variable.

\subsection{Principal Component Analysis}
While supervised learning methods are better adapted to address the classification-style of Problem \ref{ml_prob1}, techniques from the ML subfield of unsupervised learning are better suited to extracting such underlying structure in the data as desired for Problem \ref{ml_prob2}.

The first of the techniques we consider is Principal Component Analysis (PCA). As a technique, PCA extracts the most important features of a dataset through diagonalisation of the covariance matrix between the data dimensions across the dataset.
Identifying the eigenvectors of the covariance matrix and sorting them by decreasing eigenvalue, the data points can be projected onto their most significant principal components which best describe the most variance -- and hence structure -- in the data. These eigenvalues can also be reinterpreted as the explained variance, giving the amount of variance explained by each component. It is often useful to consider these explained variances as ratios, such that they have been normalised to sum to 1.

Traditional PCA, as just described, may be generalised to kernel PCA. There, the data points are conceptually mapped to a higher-dimensional space where distinguishing them becomes substantially easier due to the larger number of degrees of freedom. Then the principal components in this space can be computed, the transformed data points projected onto them, and mapped back to the original space.
However, these higher-dimensional computations are costly and can in fact be avoided altogether by using the `kernel trick'. The trick combines the above steps by defining a kernel that represents this mapping and projection, circumventing the need to actually compute in the higher-dimensional space in practice.

Whereas traditional PCA acts effectively with a linear mapping and hence linear kernel, kernel PCA can introduce non-linearity into the principal components, and hence identify non-linear structure in the data.

\begin{figure}[!t]
	\centering
	\begin{subfigure}{0.49\textwidth}
    	\centering
    	\includegraphics[width=0.98\textwidth]{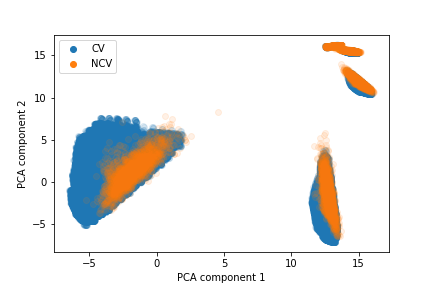}
    	\caption{$\CC[\Gr(3,12)]$ $r6$}\label{312_pcarf}
	\end{subfigure}
    \begin{subfigure}{0.49\textwidth}
    	\centering
    	\includegraphics[width=0.98\textwidth]{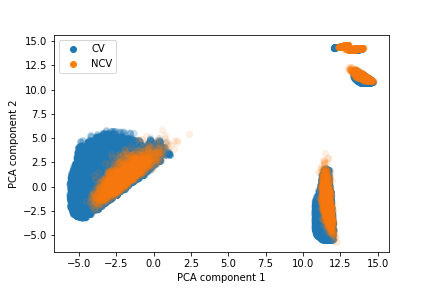}
    	\caption{$\CC[\Gr(4,10)]$ $r6$}\label{410_pcarf}
    \end{subfigure} \\
    \begin{subfigure}{0.49\textwidth}
    	\centering 
    	\includegraphics[width=0.98\textwidth]{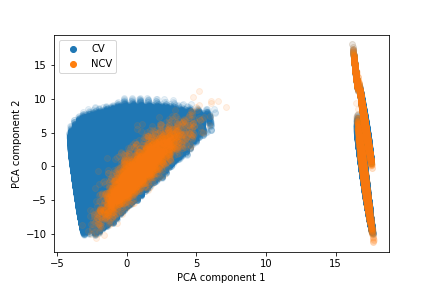}
    	\caption{$\CC[\Gr(4,12)]$ $r4$}\label{412_pcarf}
    \end{subfigure}
\caption{PCA decomposition (linear kernel) of the SSYT CV Grassmannian and NCV data for each of the respective datasets. The PCA shows that the NCV data generation is representative in the principal components.}\label{pca_rf}
\end{figure}

As a testing ground, we perform PCA (i.e. linear kernel PCA) on the Grassmannian CV data and equivalent NCV datasets. In each case the 24 explained variances over this 24-dimensional data have largest 2 normalised values 
\begin{align*}
    \CC[\Gr(3,12)] \ r6 \implies (0.592,0.138)\;,\\
    \CC[\Gr(4,10)] \ r6 \implies (0.631,0.139)\;,\\
    \CC[\Gr(4,12)] \ r4 \implies (0.488,0.179)\;,
\end{align*}
respectively. These clearly dominate the data structure (other explained variance ratios are all at least an order of magnitude smaller), and are respectively plotted as 2-dimensional plots in Figure \ref{pca_rf}. 

As shown in each of these plots the NCV data (10,000 tableaux) sits nicely within the projections of the much larger Grassmannian CV datasets. This again emphasises that the NCV data is representative in the principal components, and hence there is no significant linear structure that the supervised architectures can take advantage of in order to learn to distinguish the NCV data.
It also further supports the point that the property that distinguishes cluster from NCV SSYT is more subtle, and hence it is even more impressive that the ML methods can pick up on it so successfully.

The clear clustering of each dataset is intriguing behaviour in itself -- one we will further analyse in the following subsections. 

\subsubsection{PCA Clustering}
\begin{figure}[!t]
    \centering
    \begin{subfigure}{0.49\textwidth}
        \centering
        \includegraphics[width=0.98\textwidth]{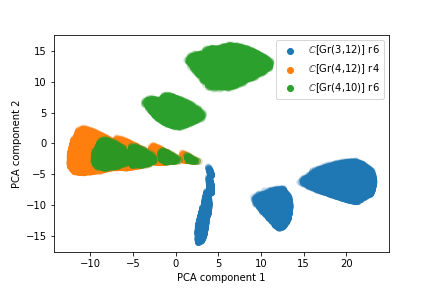}
        \caption{Linear kernel}\label{PCA_all}
    \end{subfigure}
    \begin{subfigure}{0.49\textwidth}
    	\centering
    	\includegraphics[width=0.98\textwidth]{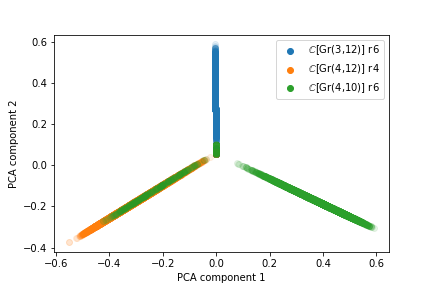}
    	\caption{Gaussian kernel}\label{KPCA_all}
    \end{subfigure}
\caption{PCA decomposition of the SSYT data for the 3 Grassmannians, using (a) linear and (b) Gaussian kernels respectively. Note there is significant overlap between $\CC[\Gr(4,10)]$ $r6$ and $\CC[\Gr(4,12)]$ $r4$ as expected, and cluster separation is largely due to padding -- hence correctly clustering according to rank. The Gaussian kernel PCA was computed over a sample of 10,000 CV SSYT from each Grassmannian due to memory limits with the full datasets.}\label{pcas_alldata}
\end{figure}

Performing PCA on all the Grassmannian CV datasets together produces an amalgamation of the aforeseen individual PCA plots for each dataset. This PCA had dominant two explained variance ratios (0.592, 0.214), again reinforcing the 2D plotting of the two most significant principal components. These two components are shown in Figure \ref{PCA_all}, and the equivalent two for a Gaussian `rbf' kernel in Figure \ref{KPCA_all}. The Gaussian kernel PCA was performed over samples of 10,000 SSYT per dataset to allow feasible computation.

The linear PCA shows a clear separation of the $\CC[\Gr(3,12)]$ $r6$ data, and a majority of the $\CC[\Gr(4,10)]$ $r6$ data separating from the $\CC[\Gr(4,12)]$ $r4$ data. This behaviour indicates simple structure differentiating the SSYT in each Grassmannian, which we may expect since the padding of the bottom row of $\CC[\Gr(3,12)]$ $r6$ data clearly separates it, whilst equally padding of the rightmost two columns for the majority of the $\CC[\Gr(4,12)]$ $r4$ data helps identify that data (where the maximum entry $>10$). 
There is overlap in the leftmost part of the linear PCA plot, where the two $k=4$ Grassmannians have common data. This was computationally confirmed to be exactly and exclusively the overlap data $\CC[\Gr(4,10)]$ $r4$. This separation supports the exceptional results in §\ref{sec:multi_clf} where the Grassmannians could be easily distinguished.

We delay the analysis of  the separation and shapes of the clusters within each Grassmannian until the next subsection. Having tried other kernels for all the Grassmannian data, the pattern appeared most striking for a Gaussian kernel. In this kernel PCA the Grassmannians are clearly separated into symmetrically distributed lines, with some overlap of the low $n$ and low rank tableaux where $k=4$, and additionally some surprising overlap with the $\CC[\Gr(3,12)]$ $r6$ data. Since the kernel was Gaussian this indicates the patches in Figure \ref{PCA_all} were approximately Gaussian distributed, which reduce to a linear parameter in each case for each dataset's line.

\subsubsection{Dissecting the Clusters}
\begin{figure}[!t]
    \centering
    \begin{subfigure}{0.49\textwidth}
        \centering
        \includegraphics[width=0.98\textwidth]{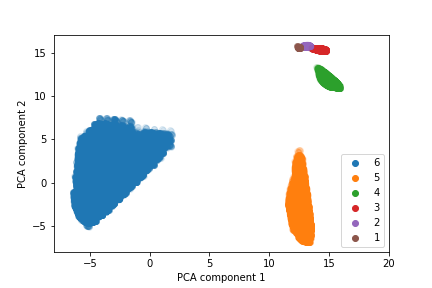}
        \caption{$\CC[\Gr(3,12)]$ partitioned according to $r \in [1,6]$}\label{312_pca_rpart}
    \end{subfigure}
    \begin{subfigure}{0.49\textwidth}
    	\centering
    	\includegraphics[width=0.98\textwidth]{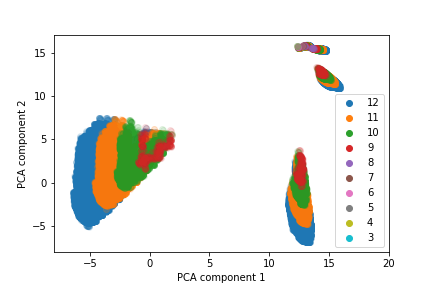}
    	\caption{$\CC[\Gr(3,n)]$ $r6$ partitioned according to $n \in [3,12]$}\label{312_pca_npart}
    \end{subfigure}
\caption{PCA decomposition (linear kernel) of the $\CC[\Gr(3,12)]$ SSYT data, plotted with partitions according to the rank $r$ or maximum entry $n$. The PCA shows that the clusters separate according to rank, whilst the differing values of $n$ expand the cluster sizes, akin to a mussel. Equivalent behaviour also holds for the other Grassmannians considered.}\label{312pca_partition}
\end{figure}

The clear separation of the individual clusters for each Grassmannian CV dataset emblematises significant data structure. The cause of this clustering should be simple, and is well shown by the  linear PCA plots for the rank-partitioning of the $\CC[\Gr(3,12)]$ data in Figure \ref{312_pca_rpart}: 
namely, each cluster corresponds to a different rank of the data,  simply identifiable by the padding, and hence easily leading to cluster separation.

The cluster relative sizes and shapes are more interesting, and are manifestly represented by the $n$-partitioning of the data in Figure \ref{312_pca_npart}, which gives the clusters a mussel-like appearance. These plots show that all the clusters exhibit higher $n \geq 9$ tableaux; however, only the smallest cluster has data for $n \leq 8$. This is likely due to the fact that as one goes to larger ranks there are more tableau boxes to fill which require a higher maximum number to satisfy the SSYT conditions. This split can be attributed to the fact that many more ranks can be used to construct tableaux when $n > 8$, as shown in Table \ref{table:number of cluster variables in Grkn}.  This table also shows why no $n \leq 8$ tableaux appear in the larger clusters, as these exclusively correspond to higher rank data.

The cluster sizes correlate with the rank; this may be expected since higher rank tableaux have more entries and thus more combinations of numbers are available. However a priori, since we know that not all SSYT are cluster variables one may not expect -- although we can build more SSYT at higher rank -- that there would also be more cluster variables; this analysis shows that this is the case.

The cluster shapes show large amounts of overlap between $n$-partitions in the data, where each tableau appears to have a counterpart with a higher $n$; one may imagine this to be due to each tableau with maximum entry (say 9) being mapped on top of an identical tableau with all the same entries except that the final largest box has entry 10, 11, or 12. As the $n$ value increases by one there is also a large number of new tableaux that can be created by increasing the largest entry to this new value, then trialing all combinations of increasing preceding entries by 1. This is what causes the clusters to grow as $n$ is increased.

This clustering behaviour is repeated in the other two datasets also, with near-identical appearance, indicative of the structure being relevant to all Grassmannian CV data, and not specifically the $\CC[\Gr(3,12)]$ data shown in Figure \ref{312pca_partition}.

Therefore we can conclude that linear PCA can be used to distinguish clusters of Grassmannian cluster variables as SSYT according to their rank and $k$ values, and gives an indication of the correlations with $n$. However, it cannot distinguish the NCV SSYT data from the CV SSYT data.

\subsection{K-Means Clustering}
An alternative unsupervised ML technique used for clustering is K-Means. This takes initialised centres for a preset number of clusters and aims to minimise the squared distance between each data point and its nearest cluster (whose sum is the inertia $\mathcal{I}$). It does this by iteratively allocating all data points to their nearest cluster, then replacing that cluster centre with the centroid of the cluster, then reallocating closest clusters for each data point. 

Clustering performance is measured with inertia $\mathcal{I}$, which is the total Euclidean squared distance between each point and its closest cluster centre. Further to the full sum, inertia may also be normalised in various ways to improve interpretability. The two normalisation methods considered were: (1) divide by the total number of points and dimensions to give the average squared distance that a datapoint was from its closest cluster centre in a single dimension $\hat{\mathcal{I}}$; and (2) divide by total number of points and dimensions, and the range of data entries $\hat{\mathcal{I}}'$ to give a more relative version of (1).

The K-Means clustering algorithm used the \texttt{sci-kit learn} standard hyperparameters \cite{sklearn}, such that 10 random initialisations are run to a convergence of 0.0001 tolerance in the inertia update or for a maximum of 300 iteration steps, with the best initialisation run selected.

To determine the optimal number of clusters to use, an elbow method is applied which reruns the clustering algorithm for a range of numbers of clusters and plots the inertia relative to the final inertia using just one cluster, and adding 0.01 $\times$ the number of clusters (so as to penalise using too many clusters). Whereas in traditional elbow methods the optimum number of clusters is represented by the largest change in gradient, in this formulation with the linear term in number of clusters added to the inertia, the lowest value across this range gives the optimal number of clusters.

\subsubsection{Distinguishing Grassmannians}
Whereas the PCA shows that the Grassmannian CV datasets can be well distinguished with simple linear structure using only a few components, we now investigate the use of K-Means on the full 24-dimensional tableaux vectors to further probe this observed clustering in the full-dimensional space.
To first exemplify the utility of K-Means, we perform clustering for a concatenated list of all the tableaux across the three datasets, with a preset number of clusters of 3.

The K-Means algorithm converges, giving inertia measures:
\begin{equation}
    \mathcal{I} = 672 000 000, \quad \hat{\mathcal{I}} = 2.32, \quad \hat{\mathcal{I}}' = 0.193,
\end{equation}
to three significant figures. Dissecting how each of the datasets split between the clusters leads to the distributions shown in Table \ref{kmeansdists_allGrs}.

\begin{table}[!t]
\centering
\begin{tabular}{|c|ccc|}
\hline
\multirow{2}{*}{Cluster} & \multicolumn{3}{c|}{Grassmannian} \\ \cline{2-4} & \multicolumn{1}{c|}{$\CC[\Gr(3,12)]$ $r6$} & \multicolumn{1}{c|}{$\CC[\Gr(4,10)]$ $r6$} & $\CC[\Gr(4,12)]$ $r4$ \\ \hline
1 & \multicolumn{1}{c|}{2656042} & \multicolumn{1}{c|}{0}       & 0       \\ \hline
2 & \multicolumn{1}{c|}{0}       & \multicolumn{1}{c|}{2951260} & 0       \\ \hline
3 & \multicolumn{1}{c|}{170}     & \multicolumn{1}{c|}{137845}  & 6346878 \\ \hline
\end{tabular}
\caption{The distributions of the three Grassmannian CV datasets between the 3 clusters generated through the K-Means process.}\label{kmeansdists_allGrs}
\end{table}

The inertia results are best interpreted using $\hat{\mathcal{I}}'$, where after clustering has converged, each datapoint is on average $<20\%$ of the range of tableaux entries away from its closest cluster centre in each dimension. Since it has been established that there is already a noticeable overlap between the datasets this clustering is quite strong, and exemplifies the power of this technique in high-dimensional clustering.

The distributions of the cluster allocations for tableaux in each dataset, as shown in Table \ref{kmeansdists_allGrs}, solidify the algorithm's ability to distinguish tableaux according to the Grassmannian they relate to. The misclassifications where tableaux in the same Grassmannian were put in a different cluster happened with proportions 0.00006, 0.045, 0 for each of the $\CC[\Gr(3,12)]$ $r6$, $\CC[\Gr(4,10)]$ $r6$, $\CC[\Gr(4,12)]$ $r4$ datasets respectively.
Further explicit analysis shows those misclassified for $\CC[\Gr(3,12)]$ $r6$ all had rank $\leq 2$ and likely the large number of zeros for these padded tableaux threw off the clustering, whilst those misclassified for $\CC[\Gr(4,10)]$ $r6$ all had rank $\leq 4$ and were \textit{exactly} the 137845 tableaux in the overlap with $\CC[\Gr(4,12)]$ $r4$.

\subsubsection{Distinguishing Cluster Tableaux}
Now using the K-Means clustering method to probe the structure differentiating SSYT which are cluster variables from those which are not, each Grassmannian CV dataset is compared against its respective NCV dataset.

Manually setting 2 clusters did not partition as well the full list of all SSYT (both cluster variables, and non-cluster variables) into their respective classes for each of the Grassmannians. Although  relatively more weight was put into one of the clusters for each case. These partitions are shown in Table \ref{kmeansdists_rf}, and show that $\sim 20\%$ of the data is misclassified under the clustering in each case.
Explicit analysis for each of the Grassmannians (where $r_{max}$ indicates the maximum rank in the dataset) shows that the misclassified NCV tableaux were all of rank $r_{max}$, which were actually \textit{all} of the NCV tableaux with rank $r_{max}$ in the NCV datasets, whilst the misclassified CV tableaux were all rank $< r_{max}$, being \textit{all} the rank $< r_{max}$ tableaux in the respective datasets.
Therefore although Table \ref{kmeansdists_rf} appears to show clustering performance related to this property, it is only an artefact of the clustering algorithm partitioning off the largest rank.

\begin{table}[!t]
\centering
\begin{tabular}{|c|cccccc|}
\hline
\multirow{3}{*}{Cluster} & \multicolumn{6}{c|}{Grassmannian} \\ \cline{2-7} & \multicolumn{2}{c|}{$\CC[\Gr(3,12)]$ $r6$} & \multicolumn{2}{c|}{$\CC[\Gr(4,10)]$ $r6$}  & \multicolumn{2}{c|}{$\CC[\Gr(4,12)]$ $r4$} \\ \cline{2-7} & \multicolumn{1}{c|}{CV}    & \multicolumn{1}{c|}{NCV}   & \multicolumn{1}{c|}{CV}    & \multicolumn{1}{c|}{NCV}   & \multicolumn{1}{c|}{CV}    & NCV   \\ \hline
1 & \multicolumn{1}{c|}{2061132} & \multicolumn{1}{c|}{595080} & \multicolumn{1}{c|}{2352760} & \multicolumn{1}{c|}{735345} & \multicolumn{1}{c|}{5963856} & 383022 \\ \hline
2 & \multicolumn{1}{c|}{1969}    & \multicolumn{1}{c|}{8031}   & \multicolumn{1}{c|}{2311}    & \multicolumn{1}{c|}{7689}   & \multicolumn{1}{c|}{3254}    & 6746   \\ \hline
\end{tabular}
\caption{The distributions of the `CV' cluster variable SSYT and the `NCV' non-cluster variable SSYT between the 2 clusters generated through the K-Means process, for each of the Grassmannian datasets respectively.}\label{kmeansdists_rf}
\end{table}

\begin{figure}[!t]
    \centering
    \includegraphics[width=0.7\textwidth]{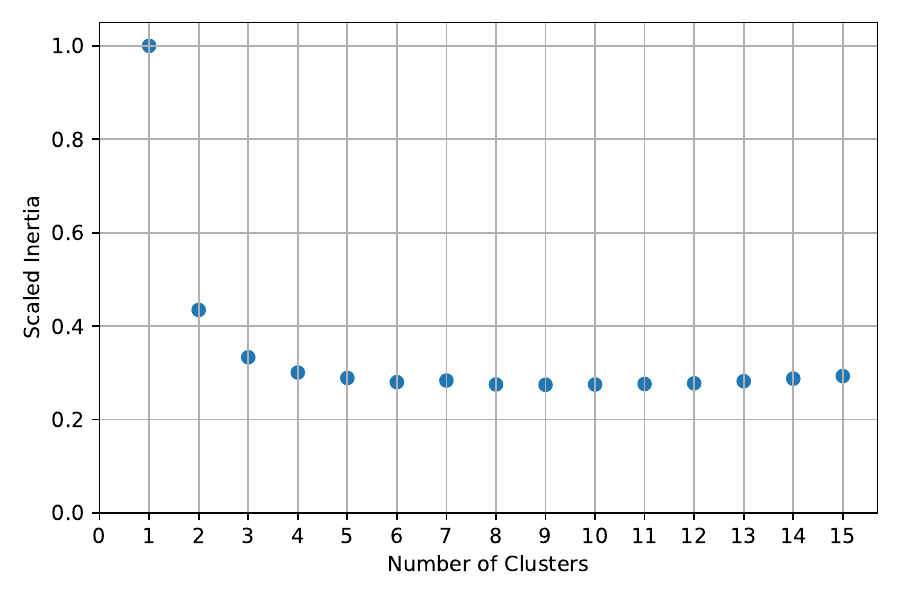}
    \caption{The elbow method for determining the optimum number of K-Means clusters when clustering the $\CC[\Gr(3,12)]$ $r6$ dataset with penalty factor of 0.01, discouraging too many clusters.}
    \label{kmeans_elbow}
\end{figure}
To further investigate this K-Means on the clustering behaviour the elbow method was applied to identify the optimum number of clusters for partitioning the CV from NCV tableaux for the $\CC[\Gr(3,12)]$ $r6$ dataset. The scaled inertia (relative to inertia with 1 cluster) is plotted for varying numbers of clusters in Figure \ref{kmeans_elbow}.

The optimum produced by this process was 9 clusters, although as can be seen from the graph there is no obvious optimum as the performance plateaus such that adding additional clusters does not improve the clustering performance.
These results clearly show that the K-Means algorithm cannot find structure in these datasets that leads to an obvious clustering, in particular one which separates the SSYT which correspond to the cluster variables.

Overall K-Means managed to distinguish Grassmannian CV datasets using the rank partitioning, corroborating the PCA results, however struggled to separate the CV and NCV tableaux, further strengthening the successes of the supervised ML methods in learning this.

\subsection{NN Gradient Saliency}
The most promising results of the ML analysis are that simple NN architectures were able to determine whether a given SSYT corresponds to a cluster variable or not.
Determining this is not possible directly, and the cluster variable and non-cluster variable datasets are certainly not distinguishable by eye.
Moreover, unsupervised methods were also unable to identify simple structure which separates these datasets, strengthening further the performance of these NNs.

\begin{figure}[!tb]
	\centering
	\begin{subfigure}{0.3\textwidth}
    	\centering
    	\includegraphics[width=0.98\textwidth]{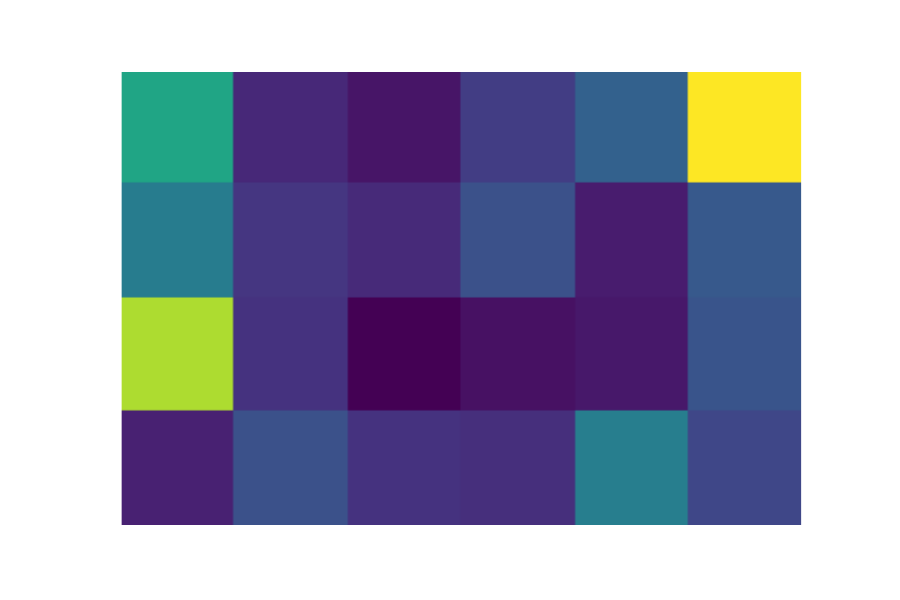}
    	\caption{$\CC[\Gr(3,12)]$, $r \leq 6$}\label{312_saliency}
	\end{subfigure}
    \begin{subfigure}{0.3\textwidth}
    	\centering
    	\includegraphics[width=0.98\textwidth]{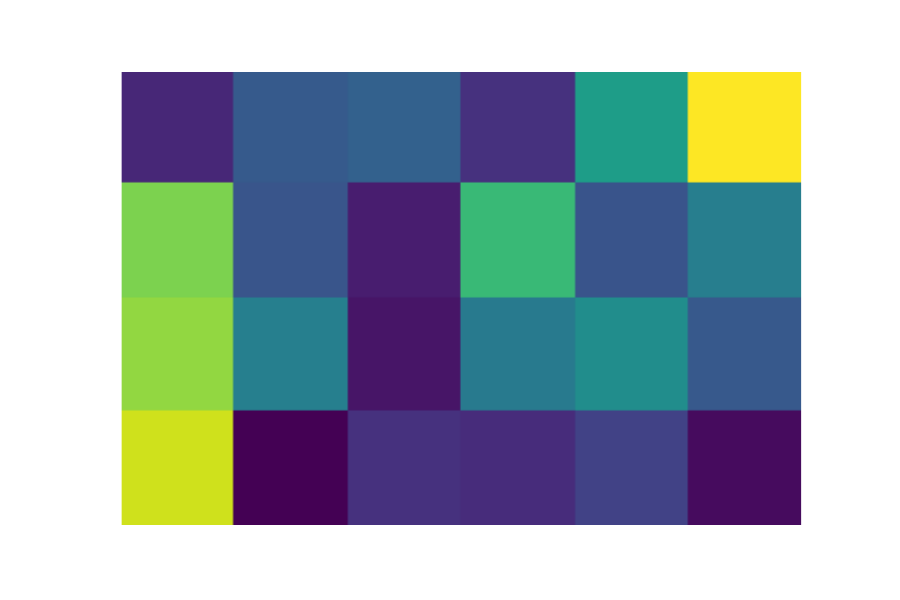}
    	\caption{$\CC[\Gr(4,10)]$, $r \leq 6$}\label{410_saliency}
    \end{subfigure} 
    \begin{subfigure}{0.3\textwidth}
    	\centering
    	\includegraphics[width=0.98\textwidth]{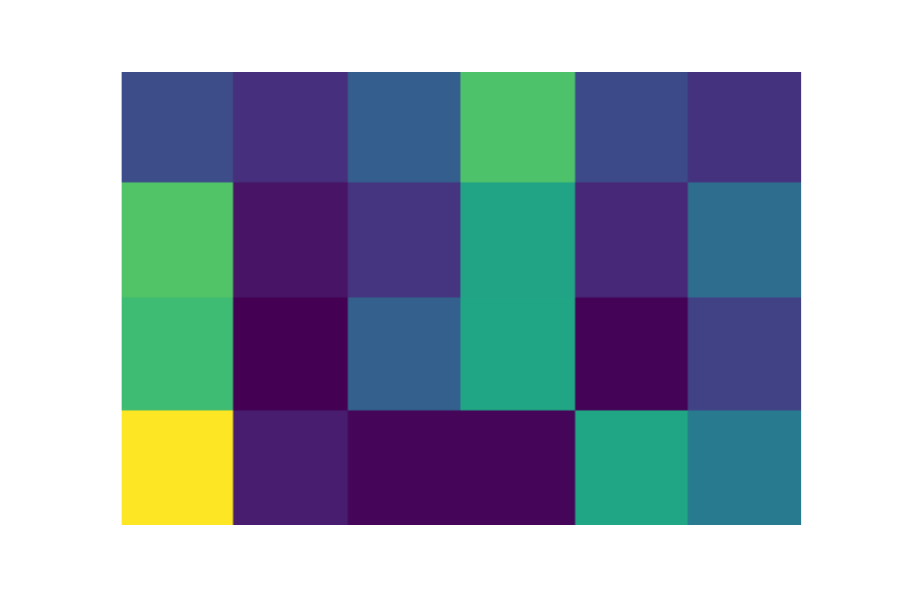}
    	\caption{$\CC[\Gr(4,12)]$, $r \leq 4$}\label{412_saliency}
    \end{subfigure}
\caption{NN gradient saliency images representing the averaged absolute values of the classification output gradients with respect to each of the respective tableaux inputs, for each of the Grassmannians considered. Lighter colours indicate the larger magnitude gradients, and hence the most dominantly useful entries for learning.}\label{saliency_images}
\end{figure}

To dissect this exceptional performance, the technique of gradient saliency was used on the trained NNs to determine which parts of the inputs most significantly contributed to the respective classification of a tableaux throughout the test dataset.
In this process, the gradient of the 0-dimensional, single entry, binary output is taken with respect to each of the input 24-dimensions, for all of the tableaux in the test dataset.
These gradients are then averaged, absolute values taken, and plotted to provide a visual representation of the more dominant features used by the NNs to perform the classification.
These images are shown for each of the binary classification investigations performed between cluster variable and non-cluster variable SSYT for each dataset in Figure \ref{saliency_images}.
Note that since higher functionality was required for the NNs to perform the saliency analysis, \texttt{tensorflow}
\cite{tensorflow2015-whitepaper} was used to construct them (with the same hyperparameters as before).

In each of the images the lighter colours indicate the gradients with the larger average absolute values over the test dataset.
Perhaps as expected for the $\CC[\Gr(3,12)]$ dataset the bottom row has no dominant features as these entries are all an artefact of padding, as is the same for the $\CC[\Gr(4,12)]$ rank 4 data where the last two columns are padded.
Interestingly though, in all cases the central columns have equivalently negligible gradients, and hence negligible effects on the learning.

The most dominant features seem to be the top-right and bottom-left entries, excluding the $\CC[\Gr(4,12)]$ rank 4 and $\CC[\Gr(3,12)]$ rank 6 padding features. Hence the structure the NNs are using to discern whether a SSYT is a cluster variable or not is likely almost entirely determined by these entries.
Some symbolic regression methods were implemented, using \texttt{gplearn}, to attempt to identify an equation that may relate these specific entries to the cluster variable prediction.
However no suitable simple equation could be found, and hence the
NNs use of these entries to determine the cluster variable nature of a generic tableaux in the Grassmannian is likely highly complicated.

\section{Conclusion}\label{conc}

In this paper, high performance computing (HPC) is applied to calculate cluster variables in Grassmannian cluster algebras $\CC[\Gr(k,n)]$. We obtained cluster variables in $\CC[\Gr(3,12)]$ up to degree $6$ (the corresponding semistandard Young tableaux has at most $6$ columns), in $\CC[\Gr(4,10)]$ up to degree $6$, and in $\CC[\Gr(4,12)]$ up to degree $4$. These cluster variables are computed for the first time and they have applications from geometry, to algebra, and to scattering amplitudes in physics \cite{ALS, DFGK, GGSVV, HP, HP2021}. Using these datasets, we verified Conjectures \ref{conj:formula of number of cluster variables} and \ref{conj:expansion of a cluster variable is a cluster variable}.

Supervised ML methods learnt to classify tableaux into each algebra with accuracy $> 0.99$, using the simple rank structure easily extractable from the data.
These architectures then also learnt to identify cluster variable SSYT from tableaux which were not cluster variables for each algebra to accuracies $\sim 0.95$. This strong performance was further supported by PCA results showing the non-cluster variable data was representative of the true cluster variable data in each case. PCA also indicated that rank was the most dominant feature explaining data variation, due to the padding structure it requires.
Clustering results with K-Means near perfectly separated the Grassmannians, but could also not differentiate the NCV tableaux, only clustering according to the rank information.

The lack of linear (and non-linear) structure in the datasets for the unsupervised methods to extract makes the supervised architecture even more impressive, confirming the utility of these advanced computational methods in analysing Grassmannian tableaux.
Through methods of NN gradient saliency the dominant tableaux features used for learning were the last non-trivial entry of the first column and first entry of the last non-trivial column, and it is likely there is structure in these entries that strongly correlates with a SSYT being a cluster variable.

\section*{Acknowledgments}
MWC would like to thank the Isaac Newton Institute for Mathematical Sciences, Cambridge, for support and hospitality during the programme `Cluster algebras and representation theory' where work on this paper was undertaken. This work was supported by EPSRC grant no EP/R014604/1.
PPD is grateful to the London Mathematical Society for grants 42035 and 42111, and to the London Institute for Mathematical Sciences for its hospitality. 
YHH would like to thank STFC for grant ST/J00037X/2.
E.~Heyes would like to thank SMCSE at City, University of London for the PhD studentship, as well as the Jersey Government for a postgraduate grant.
E.~Hirst would like to thank STFC for a PhD studentship.
JRL is supported by the Austrian Science Fund (FWF): Einzelprojekte P34602.
The computational results presented have been achieved in part using the Vienna Scientific Cluster (VSC), and the City, University of London high performance computing service (hyperion).

For the purpose of open access, the authors have applied a Creative Commons Attribution (CC BY) licence to any Author Accepted Manuscript version arising from this submission.



\bibliographystyle{elsarticle-num} 
\bibliography{references}

\begin{thebibliography}{10}
\expandafter\ifx\csname url\endcsname\relax
  \def\url#1{\texttt{#1}}\fi
\expandafter\ifx\csname urlprefix\endcsname\relax\def\urlprefix{URL }\fi
\expandafter\ifx\csname href\endcsname\relax
  \def\href#1#2{#2} \def\path#1{#1}\fi

\bibitem{FZ02}
S.~Fomin, A.~Zelevinsky, Cluster algebras {I}: foundations, Journal of the
  American Mathematical Society 15~(2) (2002) 497--529.

\bibitem{HL10}
D.~Hernandez, B.~Leclerc, Cluster algebras and quantum affine algebras, Duke
  Mathematical Journal 154~(2) (2010) 265--341.

\bibitem{JKS}
B.~T. Jensen, A.~D. King, X.~Su, A categorification of {G}rassmannian cluster
  algebras, Proceedings of the London Mathematical Society 113~(2) (2016)
  185--212.

\bibitem{ABCGPT}
N.~Arkani-Hamed, J.~Bourjaily, F.~Cachazo, A.~Goncharov, J.~Trnka,
  A.~Postnikov, {G}rassmannian geometry of scattering amplitudes, Cambridge
  University Press, 2016.

\bibitem{ALS}
N.~Arkani-Hamed, T.~Lam, M.~Spradlin, Non-perturbative geometries for planar
  {$N=4$} {SYM} amplitudes, J. High Energ. Phys.~(03) (2021) 65.
\newblock \href {https://doi.org/https://doi.org/10.1007/JHEP03(2021)065}
  {\path{doi:https://doi.org/10.1007/JHEP03(2021)065}}.

\bibitem{DDHMPS}
L.~J. Dixon, J.~Drummond, T.~Harrington, A.~J. McLeod, G.~Papathanasiou,
  S.~Marcus, Heptagons from the {Steinmann} cluster bootstrap, J. High Energ.
  Phys.~(02) (2017) 137.
\newblock \href {https://doi.org/https://doi.org/10.1007/JHEP02(2017)137}
  {\path{doi:https://doi.org/10.1007/JHEP02(2017)137}}.

\bibitem{DFGK}
J.~Drummond, J.~Foster, {\"O}.~G{\"u}rdo{\u{g}}an, C.~Kalousios, Tropical
  {G}rassmannians, cluster algebras and scattering amplitudes, Journal of High
  Energy Physics 2020~(4) (2020) 146.
\newblock \href {https://doi.org/https://doi.org/10.1007/JHEP04(2020)146}
  {\path{doi:https://doi.org/10.1007/JHEP04(2020)146}}.

\bibitem{GGSVV}
J.~Golden, A.~B. Goncharov, M.~Spradlin, C.~Vergu, A.~Volovich, Motivic
  amplitudes and cluster coordinates, Journal of High Energy Physics 2014~(1)
  (2014) 91.
\newblock \href {https://doi.org/https://doi.org/10.1007/JHEP01(2014)091}
  {\path{doi:https://doi.org/10.1007/JHEP01(2014)091}}.

\bibitem{HP}
N.~Henke, G.~Papathanasiou, How tropical are seven-and eight-particle
  amplitudes, Journal of High Energy Physics 2020~(8) (2020) 1--50.

\bibitem{Franco:2014nca}
S.~Franco, D.~Galloni, A.~Mariotti, {Bipartite Field Theories, Cluster Algebras
  and the {G}rassmannian}, J. Phys. A 47~(47) (2014) 474004.
\newblock \href {http://arxiv.org/abs/1404.3752} {\path{arXiv:1404.3752}},
  \href {https://doi.org/10.1088/1751-8113/47/47/474004}
  {\path{doi:10.1088/1751-8113/47/47/474004}}.

\bibitem{Franco:2017lpa}
S.~Franco, G.~Musiker, {Higher Cluster Categories and QFT Dualities}, Phys.
  Rev. D 98~(4) (2018) 046021.
\newblock \href {http://arxiv.org/abs/1711.01270} {\path{arXiv:1711.01270}},
  \href {https://doi.org/10.1103/PhysRevD.98.046021}
  {\path{doi:10.1103/PhysRevD.98.046021}}.

\bibitem{Franco:2003ja}
S.~Franco, A.~Hanany, Y.-H. He, P.~Kazakopoulos, {Duality walls, duality trees
  and fractional branes} (6 2003).
\newblock \href {http://arxiv.org/abs/hep-th/0306092}
  {\path{arXiv:hep-th/0306092}}.

\bibitem{CDFL}
W.~Chang, B.~Duan, C.~Fraser, J.-R. Li, Quantum affine algebras and
  {G}rassmannians, Mathematische Zeitschrift 296~(3) (2020) 1539--1583.

\bibitem{BBGL}
K.~Baur, D.~Bogdanic, A.~G. Elsener, J.-R. Li, Rigid indecomposable modules in
  grassmannian cluster categories (2020).
\newblock \href {http://arxiv.org/abs/2011.09227} {\path{arXiv:2011.09227}}.

\bibitem{Sco}
J.~S. Scott, {G}rassmannians and cluster algebras, Proceedings of the London
  Mathematical Society 92~(2) (2006) 345--380.

\bibitem{CP94}
V.~Chari, A.~Pressley, et~al., A guide to quantum groups, Cambridge University
  Press, 1995.

\bibitem{Le03}
B.~Leclerc, Imaginary vectors in the dual canonical basis of ${U}_q (n)$
  (2002).
\newblock \href {http://arxiv.org/abs/math/0202148}
  {\path{arXiv:math/0202148}}.

\bibitem{CP97}
V.~Chari, A.~Pressley, Factorization of representations of quantum affine
  algebras, Modular interfaces,(Riverside CA 1995), AMS/IP Stud. Adv. Math 4
  (1997) 33--40.

\bibitem{HP2021}
N.~Henke, G.~Papathanasiou, Singularities of eight- and nine-particle
  amplitudes from cluster algebras and tropical geometry, Journal of High
  Energy Physics 2021~(7) (2021) 1--60.

\bibitem{He:2017aed}
Y.-H. He, {Deep-Learning the Landscape} (6 2017).
\newblock \href {http://arxiv.org/abs/1706.02714} {\path{arXiv:1706.02714}}.

\bibitem{Carifio:2017bov}
J.~Carifio, J.~Halverson, D.~Krioukov, B.~D. Nelson, {Machine Learning in the
  String Landscape}, JHEP 09 (2017) 157.
\newblock \href {http://arxiv.org/abs/1707.00655} {\path{arXiv:1707.00655}},
  \href {https://doi.org/10.1007/JHEP09(2017)157}
  {\path{doi:10.1007/JHEP09(2017)157}}.

\bibitem{Krefl:2017yox}
D.~Krefl, R.-K. Seong, {Machine Learning of Calabi-Yau Volumes}, Phys. Rev. D
  96~(6) (2017) 066014.
\newblock \href {http://arxiv.org/abs/1706.03346} {\path{arXiv:1706.03346}}.

\bibitem{Ruehle:2017mzq}
F.~Ruehle, {Evolving neural networks with genetic algorithms to study the
  String Landscape}, JHEP 2017~(08) (2017) 038.
\newblock \href {http://arxiv.org/abs/1706.07024} {\path{arXiv:1706.07024}}.

\bibitem{Jejjala:2020wcc}
V.~Jejjala, D.~K. Mayorga~Pena, C.~Mishra, {Neural Network Approximations for
  Calabi-Yau Metrics} (12 2020).
\newblock \href {http://arxiv.org/abs/2012.15821} {\path{arXiv:2012.15821}}.

\bibitem{Berglund:2021ztg}
P.~Berglund, B.~Campbell, V.~Jejjala, {Machine Learning Kreuzer-Skarke
  Calabi-Yau Threefolds} (12 2021).
\newblock \href {http://arxiv.org/abs/2112.09117} {\path{arXiv:2112.09117}}.

\bibitem{Cole:2021nnt}
A.~Cole, S.~Krippendorf, A.~Schachner, G.~Shiu, {Probing the Structure of
  String Theory Vacua with Genetic Algorithms and Reinforcement Learning}, in:
  {35th Conference on Neural Information Processing Systems}, 2021.
\newblock \href {http://arxiv.org/abs/2111.11466} {\path{arXiv:2111.11466}}.

\bibitem{Arias-Tamargo:2022qgb}
G.~Arias-Tamargo, Y.-H. He, E.~Heyes, E.~Hirst, D.~Rodriguez-Gomez, {Brain webs
  for brane webs}, Phys. Lett. B 833 (2022) 137376.
\newblock \href {http://arxiv.org/abs/2202.05845} {\path{arXiv:2202.05845}},
  \href {https://doi.org/10.1016/j.physletb.2022.137376}
  {\path{doi:10.1016/j.physletb.2022.137376}}.

\bibitem{Berman:2021mcw}
D.~S. Berman, Y.-H. He, E.~Hirst, {Machine learning Calabi-Yau hypersurfaces},
  Phys. Rev. D 105~(6) (2022) 066002.
\newblock \href {http://arxiv.org/abs/2112.06350} {\path{arXiv:2112.06350}},
  \href {https://doi.org/10.1103/PhysRevD.105.066002}
  {\path{doi:10.1103/PhysRevD.105.066002}}.

\bibitem{Bao:2021ofk}
J.~Bao, Y.-H. He, E.~Hirst, J.~Hofscheier, A.~Kasprzyk, S.~Majumder, {Polytopes
  and Machine Learning} (9 2021).
\newblock \href {http://arxiv.org/abs/2109.09602} {\path{arXiv:2109.09602}}.

\bibitem{Bao:2021olg}
J.~Bao, Y.-H. He, E.~Hirst, {Neurons on Amoebae}, J. Symb. Comput. 116 (2022)
  1--38.
\newblock \href {http://arxiv.org/abs/2106.03695} {\path{arXiv:2106.03695}},
  \href {https://doi.org/10.1016/j.jsc.2022.08.021}
  {\path{doi:10.1016/j.jsc.2022.08.021}}.

\bibitem{Bao:2021auj}
J.~Bao, Y.-H. He, E.~Hirst, J.~Hofscheier, A.~Kasprzyk, S.~Majumder, {Hilbert
  series, machine learning, and applications to physics}, Phys. Lett. B 827
  (2022) 136966.
\newblock \href {http://arxiv.org/abs/2103.13436} {\path{arXiv:2103.13436}},
  \href {https://doi.org/10.1016/j.physletb.2022.136966}
  {\path{doi:10.1016/j.physletb.2022.136966}}.

\bibitem{Hirst:2022qqr}
E.~Hirst, {Machine Learning for Hilbert Series}, in: {Nankai Symposium on
  Mathematical Dialogues}: {In celebration of S.S.Chern's 110th anniversary},
  2022.
\newblock \href {http://arxiv.org/abs/2203.06073} {\path{arXiv:2203.06073}}.

\bibitem{He:2020eva}
Y.-H. He, E.~Hirst, T.~Peterken, {Machine-learning dessins
  d\textquoteright{}enfants: explorations via modular and
  Seiberg\textendash{}Witten curves}, J. Phys. A 54~(7) (2021) 075401.
\newblock \href {http://arxiv.org/abs/2004.05218} {\path{arXiv:2004.05218}},
  \href {https://doi.org/10.1088/1751-8121/abbc4f}
  {\path{doi:10.1088/1751-8121/abbc4f}}.

\bibitem{Manko:2022zfz}
M.~Manko, {An Upper Bound on the Critical Volume in a Class of Toric
  Sasaki-Einstein Manifolds} (9 2022).
\newblock \href {http://arxiv.org/abs/2209.14029} {\path{arXiv:2209.14029}}.

\bibitem{Chen:2022jwd}
S.~Chen, Y.-H. He, E.~Hirst, A.~Nestor, A.~Zahabi, {Mahler Measuring the
  Genetic Code of Amoebae} (12 2022).
\newblock \href {http://arxiv.org/abs/2212.06553} {\path{arXiv:2212.06553}}.

\bibitem{Dechant:2022ccf}
P.-P. Dechant, Y.-H. He, E.~Heyes, E.~Hirst, {Cluster Algebras: Network Science
  and Machine Learning} (3 2022).
\newblock \href {http://arxiv.org/abs/2203.13847} {\path{arXiv:2203.13847}}.

\bibitem{Bao:2020nbi}
J.~Bao, S.~Franco, Y.-H. He, E.~Hirst, G.~Musiker, Y.~Xiao, {Quiver Mutations,
  Seiberg Duality and Machine Learning}, Phys. Rev. D 102~(8) (2020) 086013.
\newblock \href {http://arxiv.org/abs/2006.10783} {\path{arXiv:2006.10783}},
  \href {https://doi.org/10.1103/PhysRevD.102.086013}
  {\path{doi:10.1103/PhysRevD.102.086013}}.

\bibitem{musiker2011compendium}
G.~Musiker, C.~Stump, A compendium on the cluster algebra and quiver package in
  {Sage} (2011).
\newblock \href {http://arxiv.org/abs/1102.4844} {\path{arXiv:1102.4844}}.

\bibitem{sklearn}
F.~Pedregosa, G.~Varoquaux, A.~Gramfort, V.~Michel, B.~Thirion, O.~Grisel,
  M.~Blondel, P.~Prettenhofer, R.~Weiss, V.~Dubourg, J.~Vanderplas, A.~Passos,
  D.~Cournapeau, M.~Brucher, M.~Perrot, E.~Duchesnay, Scikit-learn: Machine
  learning in {P}ython, Journal of Machine Learning Research 12 (2011)
  2825--2830.

\bibitem{tensorflow2015-whitepaper}
M.~Abadi, A.~Agarwal, P.~Barham, E.~Brevdo, Z.~Chen, C.~Citro, G.~S. Corrado,
  A.~Davis, J.~Dean, M.~Devin, S.~Ghemawat, I.~Goodfellow, A.~Harp, G.~Irving,
  M.~Isard, Y.~Jia, R.~Jozefowicz, L.~Kaiser, M.~Kudlur, J.~Levenberg,
  D.~Man\'{e}, R.~Monga, S.~Moore, D.~Murray, C.~Olah, M.~Schuster, J.~Shlens,
  B.~Steiner, I.~Sutskever, K.~Talwar, P.~Tucker, V.~Vanhoucke, V.~Vasudevan,
  F.~Vi\'{e}gas, O.~Vinyals, P.~Warden, M.~Wattenberg, M.~Wicke, Y.~Yu,
  X.~Zheng, \href{https://www.tensorflow.org/}{{TensorFlow}: Large-scale
  machine learning on heterogeneous systems}, software available from
  tensorflow.org (2015).
\newline\urlprefix\url{https://www.tensorflow.org/}

\end{thebibliography}
\end{document}